\documentclass[10pt,journal,compsoc]{IEEEtran}
\usepackage{cite}
\usepackage{amsmath}
\usepackage{amsfonts}
\usepackage{url}
\usepackage{graphicx}
\usepackage{subfigure}
\usepackage{caption}
\usepackage{epstopdf}
\usepackage{multirow}
\usepackage{color,xcolor}
\usepackage{cite}
\usepackage{CJK}
\usepackage{mathrsfs}
\usepackage{algorithm}
\usepackage{algorithmic}
\usepackage{multicol}
\usepackage{mathtools}
\usepackage{booktabs}
\usepackage{amssymb}
\usepackage{hyperref}
\usepackage{comment}

\begin{document}

\title{ Inferring Point Cloud Quality via Graph Similarity}
\author{Qi Yang, Zhan Ma,~\IEEEmembership{Senior Member,~IEEE}, Yiling Xu,~\IEEEmembership{Member,~IEEE}, \\ Zhu Li,~\IEEEmembership{Senior Member,~IEEE}, and Jun Sun\thanks{This paper is supported in part by National Key Research and Development Project of China Science and Technology Exchange Center (2018YFE0206700, 2018YFB1802201), National Natural Science Foundation of China (61971282, 62022038, U20A20185), and Scientific Research Plan of the Science and Technology Commission of Shanghai Municipality (18511105402). Corresponding author: Z. Ma and Y. Xu.}
\thanks{Q. Yang, Y. Xu, J. Sun are from Cooperative Medianet Innovation Center, Shanghai Jiaotong University, Shanghai, 200240, China, (e-mail: yang\_littleqi@sjtu.edu.cn, yl.xu@sjtu.edu.cn)}
\thanks{Z. Ma is from Nanjing University, Nanjing, Jiangsu, 210093, China (email: mazhan@nju.edu.cn)}
\thanks{Z. Li is from University of Missouri-Kansas City, Kansas, 64110, America (email: lizhu@umkc.edu)}
}

\IEEEtitleabstractindextext{
\begin{abstract}
     Objective quality estimation of media content plays a vital role in a wide range of applications. Though numerous metrics exist for 2D images and videos, similar metrics are missing for 3D point clouds with unstructured and non-uniformly distributed points.
	{In this paper, we propose {\sf GraphSIM}---a metric to accurately and quantitatively predict the human perception of point cloud with superimposed geometry and color impairments.}
	Human vision system  is more sensitive to the high spatial-frequency components (e.g., contours and edges), and weighs local structural variations more than individual point intensities. Motivated by this fact, we use graph signal gradient as a quality index to evaluate point cloud distortions. Specifically,  we first extract geometric {\it keypoints} by resampling the reference point cloud geometry information to form an object skeleton. Then, we construct {\it local graphs} centered at these keypoints for both reference and distorted point clouds. Next, we compute three {\it moments of color gradients} between centered keypoint and all other points in the same local graph for \textit{local significance} similarity feature. Finally, we obtain similarity index by pooling the local graph significance across all color channels and averaging across all graphs. We evaluate {\sf GraphSIM} on two large and independent point cloud assessment datasets that involve a wide range of impairments (e.g., re-sampling, compression, and additive noise). {\sf GraphSIM} provides state-of-the-art performance for all distortions with noticeable gains in predicting the subjective mean opinion score (MOS) in comparison with point-wise distance-based metrics adopted in standardized reference software.  Ablation studies further show that {\sf GraphSIM} can be generalized to various scenarios with consistent performance by adjusting its key modules and parameters.
	Models and associated materials will be made available at {\url{https://njuvision.github.io/GraphSIM}} or {\url{http://smt.sjtu.edu.cn/papers/GraphSIM}}.
\end{abstract}

\begin{IEEEkeywords}
	Objective quality assessment, human perception, graph signal processing, point cloud
\end{IEEEkeywords}}

\maketitle

\IEEEdisplaynontitleabstractindextext

%
\IEEEpeerreviewmaketitle

\IEEEraisesectionheading{\section{Introduction}\label{sec:intro}}
Point cloud has emerged as a preferred format for realistic representation of 3D objects and scenes in emerging 3D capturing and rendering technologies~\cite{schwarz2018emerging,brady2018parallel}.  A point cloud consists of a large number of unstructured 3D points with different  attributes, such as RGB color, normal, and opacity. These points are often scattered in a 3D space sparsely. A number of point cloud-based applications have emerged in recent years. Examples include resampling~\cite{chen2017fast}, enhancement~\cite{yao2010mutual,regaya20193d}, saliency detection~\cite{zheng2019pointcloud}, classification~\cite{hackel2017semantic3d,zhang2016multilevel}, segmentation~\cite{rabbani2006segmentation,tchapmi2017segcloud}, and compression~\cite{schwarz2018emerging,MPEGsys,schnabel2006octree,shao2018hybrid,gumhold2005predictive,li2020occupancy, li2020motion}. A variety of noise would inevitably be introduced by different processing stages of these applications, impairing the reconstruction quality perceived by the human visual system (HVS).

{\bf Observation.} To the best of our knowledge, most explorations for point cloud quality assessment are still applying the point-wise error measurements, such as the peak signal-to-noise ratio (PSNR). Such point-wise evaluations, however, do not accurately reflect the perception of our HVS as revealed in many pioneering assessment studies~\cite{alexiou2017subjective}. For 2D images, researchers have developed the structural similarity index (SSIM)~\cite{wang2004ssim} to better exploit the HVS characteristics when assessing the image quality. An efficient objective metric for point cloud quality evaluation is still missing. Some challenges in developing such a metric are as follows. For a given point cloud, its scattered 3D points are unstructured without explicit connections, which makes objective metrics difficult to arrange points for quantitative and effective quality measurement. It implicitly involves the superimposed subjective impacts of 3D geometry and associated attributes\footnote{In this work, we mainly emphasize on the color attributes (e.g., RGB or its variants in other color spaces).} that do not typically exist for 2D images or videos. Furthermore, point cloud processing (e.g., compression) may change the total number of points in the same point cloud, leading to issues on how to perform the fair comparison (especially using point-wise metrics). The above-mentioned challenges make quality assessment of point clouds fundamentally different from that of  2D images or videos. 

{\bf Perception.} Our work is inspired by the perception models of HVS. Our vision system  exhibits feedforward visual information extraction and aggregation from the retina (e.g., object/scene sensing) to the primary visual cortex (e.g., content understanding)~\cite{simoncelli2001natural,yamins2016using}. Light coming from an object/scene to our eyes is divided into different subbands or channels to stimulate respective neurons with nonlinear mapping in the visual cortex. Some of these channels are aggregated and some are suppressed in different layers. This processing makes
the HVS {\it frequency selective} and more sensitive to high spatial frequencies, such as the geometric structure (e.g., edge/orientation), contrast, and compound saliency~\cite{itti1998model}. Furthermore, our eyes do not directly sense the visual attributes of individual points, instead they sense local neighbor structures that are filtered by the low-pass function of our eyes optics~\cite{thibos1989image}. Thus, we can make a hypothesis that the  overall quality sensation is a weighted synthesis of individual channel components (e.g., structure and color).  It inspires us to find an efficient way to decompose the perception-related key components for effective quality assessment.

SSIM measures the image quality using the similarities within luminance, contrast, and structure components of an image. Because of the well-defined sampling for 2D image/video on a regular grid, these components can be easily derived by statistical moments of pixel distribution, such as the mean, variance, and co-variance on a  block-by-block basis~\cite{wang2004ssim}. However, the point clouds lack explicit relationships among unstructured 3D points, making it difficult to extract appropriate features for quality measurement. Dimensionality reduction, such as the 3D-to-2D projection, can be applied to inherit existing 2D image metrics for weighted quality estimation~\cite{YangPredicting2020}, but it is sensitive to projection direction and cannot characterize the 3D points distribution very well.

Recently, we have witnessed an impressive progress in {\it graph signal processing} (GSP)-based applications for point clouds~\cite{yang2013saliency,sakiyama2019eigendecomposition-free,zeng20193d}. This is mainly because the {\it graph} representation can model high-dimensional visual data (e.g., 3D point cloud) by implicitly embedding the local neighbor connections to characterize their importance. Such a mechanism correlates well with the HVS when consuming the point cloud media. Even though points are unstructured without explicit neighbor relations, our vision system implicitly applies point spread function to connect local neighbors for geometry synthesis \cite{thibos1989image}. Such point spread function can be simulated using a low-pass filter~\cite{wang2004ssim} (e.g., Gaussian kernel). Thus, we choose to utilize the GSP techniques to systematically quantify the point cloud quality.

{\bf Our Approach.} We model the point cloud quality by considering the geometry and color attributes jointly. First, we extract the keypoints of a point cloud using its geometry information, from which we construct the 3D object skeleton (e.g., contours, edges). We choose a simple yet efficient graph-based resampling method in \cite{chen2017fast} to extract keypoints.. Note that these keypoints are from the original/reference point cloud geometry, which are then used in both reference and impaired point cloud to construct local graphs. This ensures a common 3D structure for a fair comparison.
 A local graph is generated for each keypoint by setting it as the spherical center and connecting all the available neighbors within a predefined distance. We calculate the zeroth, first, and second moments of color gradients between the spherical center and its immediate neighbors in each graph (of both reference and distorted samples) to measure the local graph deformation after feature aggregation and similarity calculation. Such local graph deformation is then used to derive the graph similarity index that is consecutively pooled across different color channels (e.g., RGB, YUV or Gaussian color model (GCM)~\cite{geusebroek2001color}). After weighting all the graph similarity, we obtain the final objective scores. We call this approach {\sf GraphSIM}.

\begin{figure*}[pt]
	\centering
	\includegraphics[width=0.95\linewidth]{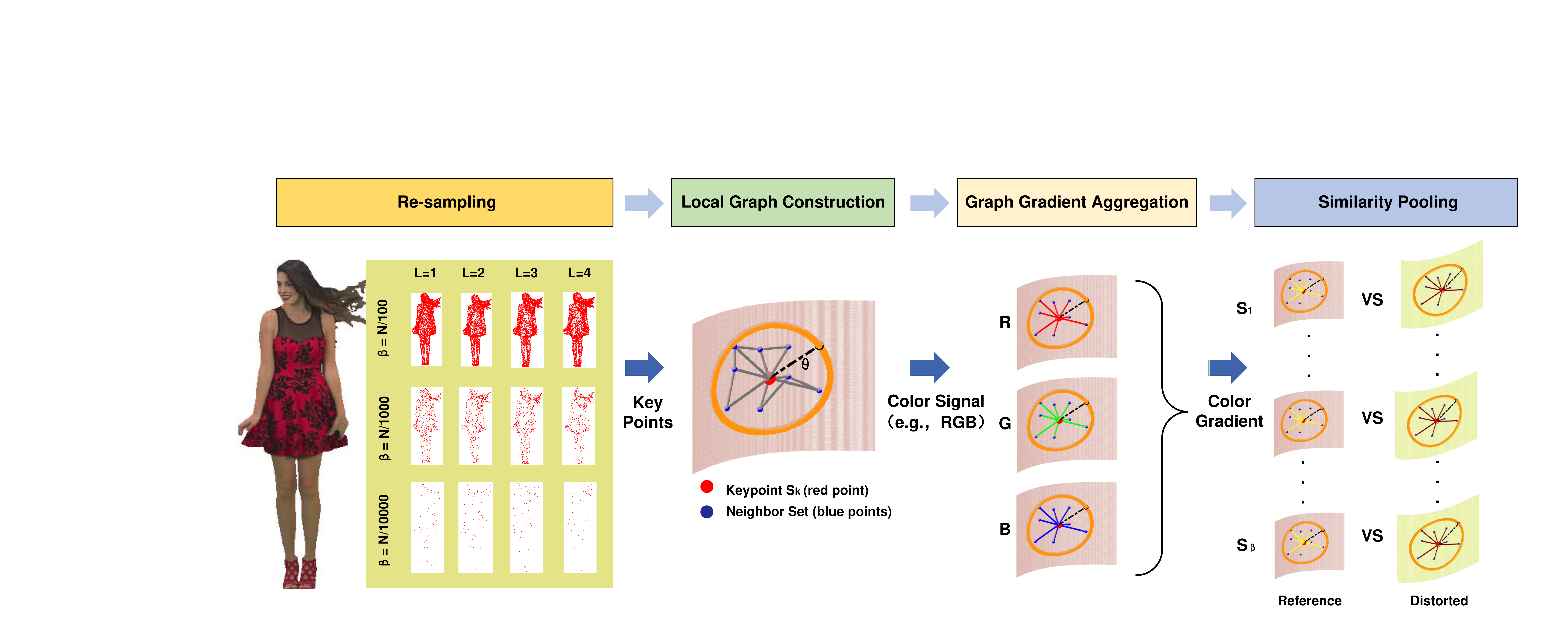}
	\caption{{\sf GraphSIM.} Objective point cloud quality assessment via GraphSIM that consists of the following steps: {\it resampling}-based geometric keypoints extraction; {\it local graph construction} centered at keypoints within clustered neighbors; {\it color gradients aggregation} for aggregating gradient moments in each graph for a specific color channel;  and {\it similarity pooling} across all color channels and local graphs. Resampling is performed using original point cloud geometry information to extract keypoints that will be used in both reference and distorted contents as common ground. Neighbor clustering assumes a sphere of radius $\theta$  (depicted as orange circle). The figure illustrates RGB color space, but other color spaces are also applicable.}
	\label{scheme}
\end{figure*}

We evaluate the {\sf GraphSIM} using two fairly large, independent and publicly accessible Point Cloud Quality Assessment databases: SJTU-PCQA~\cite{SJTUdatabase} and IRPC \cite{javaheri2019point}. The SJTU-PCQA database has 420 processed point cloud samples labeled with individual mean opinion score (MOS), and it is developed using the common test point cloud sequences suggested by the experts from Moving Picture Experts Group (MPEG) and industrial leaders. The original samples are corrupted with additive noise (e.g.,  geometry or color attribute), re-sampling, compression (e.g., octree-based) as well as their combined artifacts for assessment.
These artifacts represent the actual noise introduced in practical point cloud applications. Specifically, we use averaged Pearson linear correlation coefficient (PLCC), Spearman rank-order correlation coefficient (SROCC), and Root mean squared error (RMSE) between MOSs and objective scores to validate the performance of metrics. Our results suggest reliable and superior performance of MOS prediction for SJTU-PCQA samples with PLCC, SROCC, RMSE at 0.89, 0.88, and 1.13, respectively.

In addition, another IRPC database~\cite{javaheri2019point} is dedicated for studying the compression noise induced by the MPEG standard-compliant point cloud compression (PCC) approaches~\cite{schwarz2018emerging}, such as G-PCC (geometry-based PCC) and V-PCC (video-based PCC). The PLCC, SROCC, and RMSE are  0.94, 0.76 and 0.21 for joint {\bf People} and {\bf Inanimate} contents. This suggests the proposed {\sf GraphSIM} generalizes for  standard compliant PCC technologies.

We also examine the robustness  of {\sf GraphSIM} under different ablation settings for re-sampling mechanism, graph scale, color space, model parameters, pooling methods, and signal type.  Experiment results show  reliable generalization of proposed model in various scenarios.

{\bf Main contributions.}
\begin{itemize}
	\item To the best of our knowledge,  the proposed {\sf GraphSIM} is the {\bf first} method to assess the point cloud quality via GSP techniques, where we aggregate color gradient moments of local graphs for similarity measurement.

	\item We test {\sf GraphSIM} using independent SJTU-PCQA~\cite{SJTUdatabase} and IRPC~\cite{javaheri2019point} datasets, demonstrating the state-of-the-art accuracy in predicting the subjective MOS.
    \item We performed extensive ablation studies to show that the model generalizes to practical scenarios. Specifically, GraphSIM presents robust performance for different color spaces, re-sampling methods, graph sizes, and signal types.
\end{itemize}

The rest of this paper is organized as follow. Section~\ref{sec:related_work} reviews the studies about point cloud quality assessment. Section~\ref{sec:GSP} introduces the graph representation of point cloud as well as key graph operations for subsequent processing. Section~\ref{sec:resampling} extracts keypoints of a point cloud using the geometry information for local graph construction. Sec.~\ref{sec:GraphSIM} introduces graph construction, color gradient moments aggregation, and similarity derivation.  Section~\ref{sec:exp} presents experimental studies  to demonstrate the state-of-the-art performance of proposed model for MOS prediction. Section~\ref{sec:ablation} performs ablation studies to further demonstrate the robustness and reliable efficiency of GraphSIM.  Section~\ref{sec:conclusion} concludes the paper.

The reminder of this paper proceeds as follows: Sec.~\ref{sec:GSP} briefs the graph representation of point cloud as well as key graph operands for subsequent processing; Sec.~\ref{sec:resampling} extracts keypoints of a point cloud using the geometry information for local graph construction, color gradient moments aggregation and similarity derivation as discussed in Sec.~\ref{sec:GraphSIM}. Experimental studies are conducted in Sec.~\ref{sec:exp} to demonstrate the state-of-the-art performance of proposed model for MOS prediction, while ablation studies further evident the robustness and reliable efficiency for model generalization. Concluding remarks are drawn in Sec.~\ref{sec:conclusion}.

\section{Related Work}\label{sec:related_work}
We briefly review related work on point cloud quality assessment and modeling in this section.

A number of pioneering explorations have been made to assess the subjective point cloud quality~\cite{alexiou2017subjective,alexiou2018pointTT,alexiou2017performance,alexiou2018pointangular,alexiou2019towards,alexious2018point,alexious2018pointtowards,zhang2014subjective,alexiou2019exploiting}, from the assessment protocol,  user interaction mechanism, distortion impairment, and objective metric modeling. On the other hand, a publicly accessible SJTU-PCQA database~\cite{SJTUdatabase} has been released with 420 processed point cloud samples and associated MOSs. All of them could potentially benefit the society to study and analyze the point cloud quality.

Point-wise error measurements were first applied to evaluate the geometry distortion of point cloud, such as the  point-to-point (p2po)~\cite{Mekuria2016Evaluation}, point-to-plane (p2pl)~\cite{tian2017geometric} or point-to-mesh (p2m)~\cite{cignoni1998metro}, which could be then converted to the Hausdorff distance or MSE for geometric PSNR derivation.  Color distortion can also be measured  point-wisely  by evaluating the Y or YUV error of geometric matched point pair. All of these are adopted into the MPEG point cloud compression reference software \cite{MPEGSoft} for compression efficiency measurement. Later as analyzed extensively by EPFL lab members in their serial publications~\cite{alexiou2017subjective,alexiou2017performance,alexiou2019towards,alexious2018pointtowards,alexiou2018pointangular,alexiou2019exploiting,torlig2018novel,alexiou2019comprehensive}, these point-wise distance based metrics are not well correlated with the subjective assessments, and presented unreliable prediction accuracy. \cite{su2019perceptual,zerman2019subjective} also tested the performance of these metrics under the distortion caused by typical compression methods, such as MPEG Point cloud Test Model Category 2 (TMC2)\cite{MPEGTMC2} and reached the same conclusion.

Motivated by the projection-based approach used for MPEG point cloud compression, Alexiou {\it et al.}~\cite{alexiou2019exploiting} recently proposed to project 3D point cloud to 2D planes, by which classical image objective quality metrics (e.g., SSIM~\cite{wang2004ssim}) could be applied. Experiments showed better efficiency under certain types of impairments, but more deep investigations were highly desired for reliable and consistent quality prediction. Yang {\it et al.}\cite{YangPredicting2020} proposed to use six-sides-projection to collect texture and depth maps, and then fused the features extracted from these maps as point cloud quality index. Experiment results showed that the injection of depth map significantly improves the evaluation accuracy. However, the projection based methods are sensitive to projection direction and number of projection planes, which limits the development of this branch.

\section{Point Cloud via Graph Representation}\label{sec:GSP}
We briefly introduce key concepts of graph signal processing applied in this paper for re-sampling and local graph construction. Additional details on graph signal processing can be found in~\cite{shuman2013emerging}.

\subsection{Graph} Suppose a point cloud ${\vec{\bf P}}$ has $N$ points and each point has $K$ attributes given as
$\vec{\bf P}$= [$\vec{X}_1, \vec{X}_2,\ldots,\vec{X}_N$]$^T\in\mathbb{R}^{N\times K}$. In this work, we consider 3D geometry of points (i.e., ($x$, $y$, $z$) coordinates) and three-dimensional color attributes (in RGB), which leads to the tuple representation of the $i$-th point $\vec{X}_i$= ($x_i,y_i,z_i,R_i,G_i,B_i$). We further denote separate geometry and color channels as $\vec{X}^O_i$ = ($x_i, y_i, z_i$) and $\vec{X}^I_i$=($R_i, G_i, B_i$),  respectively\footnote{Note that superscript ``O'' stands for geometric {\it occupancy}, and ``I'' is for color {\it intensity}.}, i.e., $\vec{X}_i$=[$\vec{X}^O_i,\vec{X}^I_i$].  $N$ is usually a large number; for example, $N$ = 729,133 for MPEG point cloud  ``RedandBlack''.

Based on the fact that our optic vision can be simulated using a low-pass filter~\cite{wang2004ssim}, e.g., Gaussian model, we propose to construct the graph representation of a point cloud by encoding the {\it local neighbor connection weights} into an {\bf adjacency matrix} $\mathbf{W}\in\mathbb{R}^{N\times N}$. Each effective connection between two points having positive weight is referred to as ``edge''. We can call the ``point'' as ``vertex'' following the convention in graph theory.
We formulate the edge/connection weight between $\vec{X}_i$ and  $\vec{X}_j$ using the geometric distance as
\begin{equation}\label{adjancy matrix}
\mathbf{W}_{\vec{X}_i,\vec{X}_j}=
\begin{cases}
e^{-\frac{\|\vec{X}^O_{i}-\vec{X}^O_{j}\|^{2}_{2}}{\sigma^2}}& \text{if } \|\vec{X}^O_{i}-\vec{X}^O_{j}\|^{2}_{2}\leq\tau\\
0& \text{otherwise.}
\end{cases}
\end{equation}
where $\vec{X}^O_i, \vec{X}^O_j$ are 3D coordinates of $\vec{X}_i$ and $\vec{X}_j$, $\sigma$ represents the variance of graph nodes and $\tau$ is the Euclidean distance threshold used for clustering neighbor points into the same graph. Finally, a point cloud $\vec{\bf P}$ can be represented using graphs involving points and their neighbor connections,  noted as $\vec{\bf G}_{\vec{\bf P}, \mathbf{W}\left(\vec{\bf P}\right)}$.

\subsection{Operand}
We first introduce a diagonal degree matrix $\mathbf{D}$ used for measuring the edge density of each vertex.
For $\vec{X}_i$, its connection density is $d_{i}=\sum_{\vec{X}_j} \mathbf{W}_{\vec{X}_i,\vec{X}_j}$.
Overall, we have $\mathbf{D}= {\sf diag}(d_{1},...,d_{N})\in \mathbb{R}^{N \times N}.$

A signal ${f}$ can be defined at each graph vertex using a vector. One example of $f$ at $\vec{X}_i$ is the vectorized color intensities: $f(\vec{X}_i) = [R_i, G_i, B_i]$.  Given an edge between $\vec{X}_i$ and $\vec{X}_j$, denoted as  $e=(\vec{X}_i,\vec{X}_j)$,  its {\it graph edge derivative} of  ${f}$ can be derived from graph Laplacian regularizer\cite{bai2018graph},
\begin{align}
\frac{\partial {f}}{\partial e}\Big|_{\vec{X}_i}:={\sqrt{\mathbf{W}_{\vec{X}_i,\vec{X}_j}}}[f(\vec{X}_i)-f(\vec{X}_j)],\label{eq:original_gradient}
\end{align} and thus corresponding {\it graph gradient} of $f$ is the collection of all edge derivatives connected to  $\vec{X}_i$ as
\begin{equation}
\nabla_{\vec{X}_i} {f}=
\sum_{e\in E}\frac{\partial {f}}{\partial e}\Big|_{\vec{X}_i}, \label{gradent}
\end{equation}
with $E$ as a set of edges connected to $\vec{X}_i$.

It then leads to the {\bf Graph Laplacian matrix},
\begin{align}
    \mathbf{L}=\mathbf{D}-\mathbf{W},
\end{align}
 which is a {\it difference operand} on graphs. For any signal ${f}\in \mathbb{R}^{N}$, its Laplacian operation is
\begin{equation}
(\mathbf{L}f)_{i}=\sum\nolimits_{\vec{X}_j\in \mathcal{N}_i} \mathbf{W}_{\vec{X}_i,\vec{X}_j}\cdot\left[f(\vec{X}_i)-f(\vec{X}_j)\right], \label{lap}
\end{equation} where $\mathcal{N}_i$ is a set of neighbors attached to $\vec{X}_i$, and $f$ can be the RGB color intensity,  normal or other attributes as well.


\section{Point Cloud Re-sampling} \label{sec:resampling}
As suggested in~\cite{wang2004ssim} and other neuroscience developments, our HVS weighs more to the structural information of perceived object or scene. In image quality assessment, structural features (e.g., high-frequency edge and contours) have been widely adopted in many popular objective quality  models\cite{xue2013gradient,ni2017esim,fu2018screen,ni2018gabor,yang2019modeling}. The similar methodologies have been also extended to 3D objects \cite{huang2013edge,meinhardt20093d}. For a 3D object, our vision system would first capture the general 3D structure, rather individual point intensity. Such 3D structure is the discriminative {\it geometric skeleton} (e.g., edges, contours) of the object. Corresponding points with this skeleton are referred to as the ``geometric keypoints''.

We could obtain these keypoints via point cloud re-sampling.
These keypoints should form the edges, contours, and skeleton for quality assessment, which mostly belong to the high spatial-frequency band that is sensitive to the HVS. We therefore choose a simple yet efficient high-pass graph filtering method in \cite{chen2017fast} to fulfill this task. Graph filtering is briefed below. Please refer to \cite{chen2017fast} for more details.

Let $A\in \mathbb{R}^{N\times N}$ be a {\it graph shift operator}, which can be formulated using the adjacency matrix $\mathbf{W}$, or  transition matrix $\mathbf{D}^{-1}\mathbf{W}$, or graph Laplacian matrix $\mathbf{L}$. A linear, shift-invariant graph filter is a polynomial function of $A$,
\begin{equation}\label{filter}
h(A)=\sum\nolimits^{L-1}_{l=0}h_{l}A^{l}=h_0I+h_1A+...+h_{L-1}A^{L-1},
\end{equation}
where $h_l$ is $l$-th coefficient, and $L$ is the length of graph filter.
For $h(A)$, a Haar-like graph filter is selected to implement the high-pass filtering,
\begin{equation}\label{high-pass}
\begin{aligned}
h_{HH}(A)=&I-A\\
=&V\begin{bmatrix} 1-\lambda_1&0&...&0\\
0&1-\lambda_2&...&0\\
\vdots&\vdots&\ddots&\vdots\\
0&0&\ldots&1-\lambda_N
\end{bmatrix}V^{-1}.
\end{aligned}
\end{equation}
Here, $A=\mathbf{D}^{-1}\mathbf{W}$, $\lambda_i$ and $V$ are the eigenvalues, eigenvectors of $A$. Thus, the frequency response of $\vec{X}_i$ is
\begin{align}
F(\vec{X}_i) = h_{HH}(A)\cdot \vec{X}_i = \vec{X}_i-\sum\nolimits_{\vec{X}_j\in \mathcal{N}_i}A_{\vec{X}_i,\vec{X}_j}\cdot \vec{X}_j, \label{eq:freq_response}
\end{align} which is then utilized to order the points in spatial frequency domain for sampling.

Applying re-sampled high-frequency keypoints for subsequent quality assessment not only fits the perceptual intuition, but also significantly reduce the computational complexity which is of practical interests in applications. Other resampling methods can be applied as well in {\sf GraphSIM}. More discussions are given in Sec.~\ref{sec:ablation}.

Note that geometric keypoints are extracted from the reference point cloud only using its geometry information. They are then leveraged to construct local graphs for both reference and distorted point clouds. This ensures the fair comparison with the common 3D geometric structure.

\section{{\sf GraphSIM}: Measuring Point Cloud Quality via Graph Similarity}\label{sec:GraphSIM}
This section details the development of {\sf GraphSIM} for quantitatively point cloud quality estimation, in which keypoints resampling, local graph construction, color gradients aggregation, and similarity derivation are involved.

\subsection{Keypoints Resampling}
Given a reference point cloud $\vec{\bf P}_r$, we first re-sample it to derive the geometric keypoints,
\begin{equation}\label{fast_gen}
\vec{\bf P}_s=\lfloor\Psi(\vec{\bf P}_r, L)\rfloor_{\beta} \in \mathbb{R}^{\beta\times6}, \beta\ll N,
\end{equation}
where $\Psi(\cdot)$ is a high-pass graph filter suggested in Sec.~\ref{sec:resampling}, $L$ is filter length, and  $\beta$ is the number of effective keypoints after re-sampling. $\vec{\bf P}_s$ represents re-sampled points from the original point cloud $\vec{\bf P}_r$, and $\lfloor\cdot\rfloor_{\beta}$ is referred to as the filter-based sampling operation that sets $\beta$ as the total number of output points to sample the frequency-ordered points. Typically, the total number of output points is less than $N$.

Specifically, each point is associated with its corresponding weighting score according to the frequency response derived using Eq.~\eqref{eq:freq_response}, e.g., a higher frequency point often comes with a larger score~\cite{chen2017fast}. This weighting score is then used as the probability measurement to decide whether associated point can be selected as the keypoint. Those high-frequency points that are often with larger scores are more likely to be chosen as keypoints.
We show results under different $L$s and $\beta$s in Fig. \ref{scheme}. It reveals that graph-based resampling filter tries to retain points close to the object contour, edges (e.g., hair, skirt hemlines and facial features) along with the decrease of $\beta$. Specifically, the plot having $L=4$ and $\beta = N/1000$  shows the best results, yielding discriminative shirt's hemline and face subjectively.
On the other hand, as reported in~\cite{chen2017fast}, the bigger $L$ comes with the larger reception field, better filtering performance, and higher complexity.
Therefore, we choose $\beta = N/1000$, and $L=4$ for subsequent steps to well balance the complexity and efficiency. Other resampling methods and $\beta$ selection will be further explored in Sec.~\ref{sec:ablation}.

\subsection{Local Graph Construction} Each keypoint in $\vec{\bf P}_s$ is used as the center to construct {\it local graph} in both reference $\vec{\bf P}_r$ and distorted point clouds $\vec{\bf P}_d$. For $k$-th keypoint $\vec{s}_k$ in $\vec{\bf P}_s$, we cluster its neighbors using the Euclidean distance of corresponding geometry components in both $\vec{\bf P}_r$ and  $\vec{\bf P}_d$, i.e.,
\begin{equation}\label{local set}
\begin{aligned}
\vec{\bf V}_{(\vec{\bf X}_r, \vec{s}_k)} &= \{\vec{X}_r\}\subset \vec{\bf P}_r, \|\vec{X}_r^{O}-\vec{s}_{k}^{O}\|^{2}_{2}\leq\theta,\\
\vec{\bf V}_{(\vec{\bf X}_d, \vec{s}_k)}&= \{\vec{X}_d\}\subset \vec{\bf P}_d, \|\vec{X}_d^{O}-\vec{s}_{k}^{O}\|^{2}_{2}\leq\theta.
\end{aligned}
\end{equation} $\vec{\bf V}_{(\vec{\bf X}_r, \vec{s}_k)}$ and $\vec{\bf V}_{(\vec{\bf X}_d, \vec{s}_k)}$ represent the groups of neighbors of $\vec{s}_{k}$ in reference and distorted content. 
We then follow (\ref{adjancy matrix}) to construct connected local graphs with $\vec{s}_{k}$ as the spherical center. In general, the selection of $\tau$ in \eqref{adjancy matrix} depends on $\theta$, and $\sigma$  is a function of $\tau$ used for the adjacency matrix $\mathbf{W}$s and corresponding graphs.

\begin{figure*}[t]
	\centering
	\includegraphics[width=0.95\linewidth]{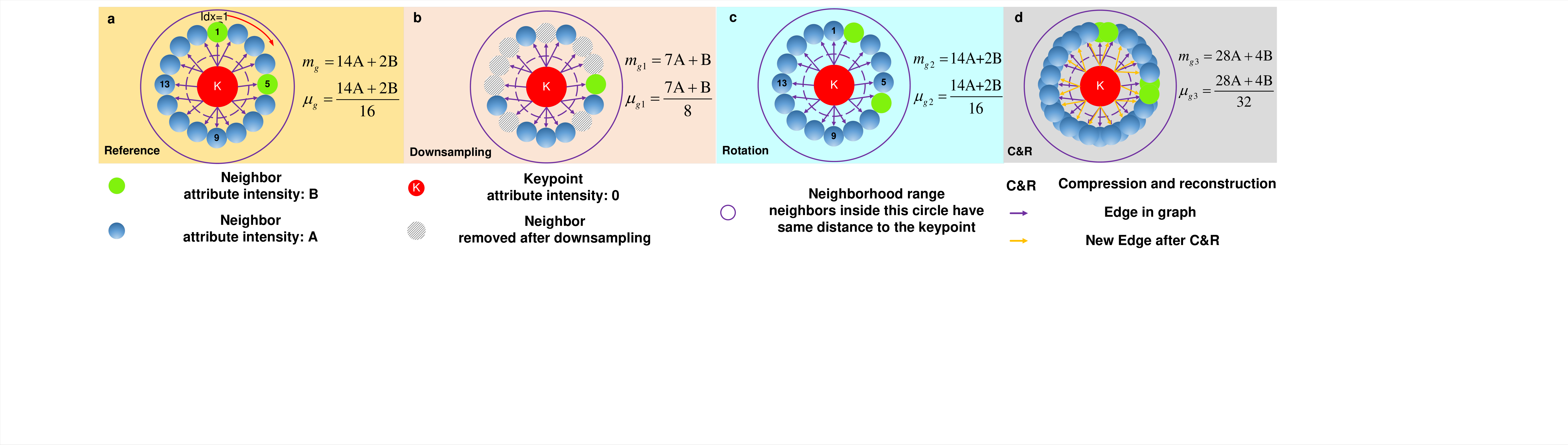}
	\caption{{\bf A Toy Example}. a) a local graph as the reference; b) reconstructed graph with ``downsampling'' artifacts in which partial points are removed from the reference; c) reconstructed graph after performing the clock-wisely geometric rotation from the reference; d) reconstructed graph with Compression and Reconstruction (C\&R) distortions in which additional points are generated in processing from the reference. These newly generated points are connected to the keypoint using yellow arrows. For the sake of simplicity, we assume all neighbors in purple rings have the same distance to the keypoint, and we set the local neighbor connection weight as 1.}
	\label{fig:feature}
\end{figure*}

\subsection{Color Gradient Features}
We first use zeroth, first and second moments of color gradients to respectively represent the mass, mean and variance features, by which we try to well illustrate the color distribution in a local graph. As will be shown in later discussion, each feature has its unique effect in distortion measurement. For the sake of simplicity, we use  a one-dimensional single-channel signal, i.e., $f \in R^{N\times1}$ for illustration. $N$ represents the number of vertices in the graph. Channel-wise pooling of multidimensional signal will be unfolded in next section.
\subsubsection{Zeroth Moment: Gradient Mass $m_g$}

Existing studies have reported that the HVS is extremely sensitive to the texture gradient for perceiving a 3D surface~\cite{tsutsui2002neural,li2000perception}. In the meantime, Curcio {\it et al.}~\cite{curcio1990human} and Guo {\it et al.} ~\cite{guo2017modeling} have also suggested that the sensitivity of HVS is non-uniform. This is because of the unequal distributed cone and rod cells on retina. Such non-uniform perceptual sensation can be well characterized using a generalized Gaussian model with respect to the eccentricity. Thus, we propose to measure the perceptual impact using the distance-weighted graph signal gradients.

More specifically, we use (\ref{gradent}) to derive the color gradients for vertex $\vec{s}_k$ in local graph, which are summed up for the zeroth moment based gradient mass  $m_g$ measurement,
\begin{align}
m_g&=\bigtriangledown_{\vec{s}_k} {f}= \mathbf{SUM}(\frac{\partial {f}}{\partial e}\Big|_{\vec{s}_k}) \nonumber\\
&=\sum\nolimits_{\vec{X}_j\in\mathcal{N}_k}\sqrt{\mathbf{W}_{\vec{X}_j,\vec{s}_k}}[f(\vec{X}_j)-f(\vec{s}_k)].
\label{local gradent}
\end{align}
The $\mathbf{SUM}(\cdot)$ is for element-wise sum. As aforementioned, graph vertex  $\vec{s}_k$ corresponds to a specific keypoint of a sampled point cloud, which is the center of a constructed graph. Thus, \eqref{local gradent} can be utilized to derive the $m_g$ for each graph against its center keypoint. Note that we only consider the points that have an effective connection with $\vec{s}_k$, i.e., $\forall \vec{X}_j \in \mathcal{N}_k, \mathbf{W}_{\vec{X}_j,\vec{s}_k}\neq 0$.
For a specific neighbor point, its contribution to the overall graph gradient mass $m_g$ is jointly determined by the Euclidean distance-based weight $\mathbf{W}_{\vec{X}_j,\vec{s}_k}$ and the attribute gradient $[f(\vec{X}_j)-f(\vec{s}_k)]$. 

For a 2D image, relative changes of local pixel intensity can be easily captured by the vision system, reflecting the quality sensation distortion.  Such change can be quantitatively measured using local contrast, such as the variance-based contrast used in SSIM~\cite{wang2004ssim}.  Similarly, we calculate the color gradient in \eqref{local gradent} to represent such local color intensity variations. It is then weighted by the Euclidean distance factor to jointly consider the superimposed impairments of geometric and colormetric attributes. Note that a similar function as (\ref{local gradent}) is used in~\cite{bai2018graph} as weighted graph total variation for image denoising.



The $m_g$ can reflect {\it point density} variations if we assume the unit graph (or after normalization). This is devised to deal with the scenario that the number of effective points in a graph changes due to the injection of impairments, such as downsampling, compression. By comparing the impaired graph in Fig.~\ref{fig:feature}(b) and reference graph in Fig.~\ref{fig:feature}(a), $m_{g1}$ is halved from the reference $m_{g}$, e.g., $m_{g1}=0.5*m_{g}$. The difference of $m_g$  can be used to measure the downsampling induced point loss. The  $m_g$ term, however,  is barely utilized in metrics for 2D image/video. This is mainly because pixels in image or video blocks are usually impaired with intensity variations with the same  number of appearance. On the contrary, points may emerge or vanish in point cloud due to a variety of processing computations.

Nevertheless,  $m_g$ is not capable of efficiently handling the geometric displacement, perceptual inconsistency, etc. For instance, for a local graph, its  $m_g$ will be the same if the point locations change due to motion displacement (e.g., rotation), but point intensities are retained. An example is illustrated in Fig. \ref{fig:feature}(c), where $m_{g2} = m_g$ though rotation clearly changes the geometric appearance.   On the other hand, as observed in subsequent standard compliant PCC approaches, point density is improved after reconstruction (e.g., see Fig. \ref{fig:feature}(d)), but perceptual sensation is almost the same
as the ground truth.  This also leads to the inconsistency with the  $m_g$ measurement. Thus, we further introduce the first and second moments of graph color gradients to improve the quality measurement.
\subsubsection{First Moment: Gradient Mean $\mu_g$}
 We first extend  $m_g$ to derive the mean by simple normalization:
\begin{align}
\mu_{g} =\frac{1}{N}(\bigtriangledown_{\vec{s}_k} {f}), \label{gr}
\end{align}
where $N=|\mathcal{N}_k|$ is the total number of points connected with the keypoint in a graph.
 It supplements the $\mu_g$ to infer the quality if $m_g$ cannot give sufficient discriminative difference for evaluating the perceptual sensation.

The $\mu_{g}$ measures the averaged color gradient difference between the keypoint and its neighbors in a specific local graph, revealing the averaged local contrast variations. As exemplified in Fig. \ref{fig:feature}(d), the process of Compression and Reconstruction (C\&R) would significantly increase even double the number of points (e.g., Table \ref{tab:appendix}), leading to $m_{g3} = 2*m_{g}$.  But the perceptual sensation is almost the same, having the calculation of $m_g$ conflict with the subjective assessment. Therefore, we normalize the $m_g$ with the number of points as another supplementary measurement. In this example, $\mu_{g3}=\mu_{g}$.



\subsubsection{Second Moment: Gradient Variance and Co-variance}
 In Fig. \ref{fig:feature}(c), we have given another example having the geometric rotation. Specifically, when the neighbors are rotated clockwisely by a certain degree, the gradient mass and gradient mean are the same, but two graphs, e.g.,  Fig. \ref{fig:feature}(a) and (c), present different geometrical pattern for perception. In this case, neither zeroth nor first moment can handle this type of impairment because $m_g$ and $\mu_g$ do not consider any motion induced variations.  The neighbors with the same signal attribute and Euclidean distance to the keypoint play equal importance role in zeroth and first moment features, such as the green points in the reference graph of Fig. \ref{fig:feature}(a).    Often times, the second moment feature, e.g., variance or co-variance can be used to measures the relative variation trend of signal attributes, thus, it could be a potential metric for measuring such distortions. Thus, we propose to apply the second moment feature for this purpose.

In order to calculate gradient variance and co-variance, point matching is first performed by using the point-wise Euclidean distance, and is used to assure that corresponding points are from the same geometrical location in each graph, e.g., the green point with $\tt idx=1$ in Fig. \ref{fig:feature}(a) and the blue point with $\tt idx=1$ in Fig. \ref{fig:feature}(c).
 Such point-wise matching is also used in existing metrics~\cite{schwarz2018emerging} for fair and quantitative comparison.

 For two graphs centered at $\vec{s}_k$, e.g., reference graph $G_{\vec{\bf r},\vec{s}_k}$ and distorted graph $G_{\vec{\bf d},\vec{s}_k}$, we choose the one having the less points as the baseline, and then regulate another one by point matching to guarantee the same number of points. We utilize the nearest distance search to perform the point matching.

 More specifically, we assume the location index $\tt idx$ of the first green point as 1, e.g., $\tt idx$ = 1, and the $\tt idx$ increases clockwisely. The same indexing pattern is applied to other impaired examples. After injecting the rotation in Fig.~\ref{fig:feature}(c), the position with $\tt idx$ =1 is occupied by another point, e.g., the blue one instead of original green one.   We then could have the same length point vector for variance and co-variance derivation after performing the nearest neighbor search based point matching.

 For simplicity, we use  $\tilde{G}_{\vec{\bf r},\vec{s}_k}$ and $\tilde{G}_{\vec{\bf d},\vec{s}_k}$ to represent the queues after point matching. Subsequently, we derive the variance and co-variance upon these two queues having same number of scattered points.




Following the gradient calculation, we have the edge weighted gradient for $\vec{X}_j$ as
\begin{align}
g_{\vec{X}_j,\vec{s}_k} = \sqrt{\mathbf{W}_{\vec{X}_j,\vec{s}_k}}\cdot(f(\vec{X}_j)-f(\vec{s}_k)).
\label{gradient_seq}
\end{align} It comprises the weighted gradient distribution of all connected points for a specific graph, as
\begin{align}
{\vec{g}}_{\vec{s}_k}:=
&\left[\left\{ g_j = g_{\vec{X}_j,\vec{s}_k} \right\}, s.t.\mbox{~~}  \vec{X}_j\in \mathcal{\tilde{N}}_{\vec{s}_k}\right]. \label{weighted_grad}
\end{align} We use $g_j$ for simplicity, and $\mathcal{\tilde{N}}_{\vec{s}_k}$ is for either $\tilde{G}_{\vec{\bf r},\vec{s}_k}$ or  $\tilde{G}_{\vec{\bf d},\vec{s}_k}$.


It then leads to the variance derivation of edge weighted gradients as,
\begin{align}
\sigma^2_g := \sigma_{\vec{g}_{\vec{s}_k}}^2=\frac{\sum(g_j-\bar{g})^2}{N},
\label{variance}
\end{align}
where $g_j$ represents $j$-th element in ${\vec{g}}_{\vec{s}_k}$, $\bar{g}$ represents averaged gradient of $\vec{g}_{\vec{s}_k}$, and $N=|\mathcal{\tilde{N}}_{\vec{s}_k}|$. Similarity, we calculate the co-variance as
\begin{align}
c_{\vec{g}_{\vec{s}_k}\vec{g}'_{\vec{s}_k}}= {\sf cov}({\vec{g}}_{\vec{s}_k},{\vec{g}}'_{\vec{s}_k})
= {\sf E}[{\vec{g}}_{\vec{s}_k}\cdot{\vec{g}}'_{\vec{s}_k}]- {\sf E}[{\vec{g}}_{\vec{s}_k}]\cdot {\sf E}[{\vec{g}}'_{\vec{s}_k}],
\label{gcf}
\end{align} with ${\vec{g}}_{\vec{s}_k}$ and ${\vec{g}}'_{\vec{s}_k}$ representing the weighted gradient distribution of respective reference and impaired point clouds. For simplicity, we note $c_{\vec{g}_{\vec{s}_k}\vec{g}'_{\vec{s}_k}}$ as $c_g$.



\subsection{{\sf GraphSIM}}


For each color channel, we have mass $m_g$, mean $\mu_g$ and variances $c_g$ derived by the statistical movements of the gradient distribution for both graphs from the reference and impaired point clouds. These features represent the local graph significance that can be utilized for quality measurement quantitatively. We then propose to fuse these three feature-based similarities together to have a general index for color channel $C$, i.e.,
\begin{align}
\label{feature_pooling}
S_{{\vec{s}_k},C}&= {\sf SIM}_{m_g}\cdot{\sf SIM}_{\mu_g}\cdot {\sf SIM}_{c_g},
\end{align} where
\begin{align}
\mathsf{SIM}_{m_g} &=
\frac{2m^r_g\cdot m_g^d+T_0}{ (m_g^r)^2+ (m_g^d)^2+T_0},\\
\mathsf{SIM}_{\mu_g} &=
\frac{2\mu^r_g\cdot \mu_g^d+T_1}{ (\mu_g^r)^2+ (\mu_g^d)^2+T_1}, \\
{\sf SIM}_{c_g}&=\frac{c_g+T_2}{\sigma^r_{{{g}}}\cdot\sigma^d_{{{g}}}+T_2}.
\end{align} $T_0$, $T_1$ and $T_2$ are  small no-zero constants to prevent numerical instability. Note that ${\sf SIM}_{m_g}$, ${\sf SIM}_{\mu_g}$ and ${\sf SIM}_{c_g}$ are in the range of [-1,1].  We use the superscripts $r$ and $d$ to indicate the reference and distorted samples, respectively. Note that similar channel-wise pooling strategy has been widely used, such as in well-known SSIM metric~\cite{wang2004ssim}.



\begin{table}[t]
	\centering
	\caption{Sample Point Clouds Illustration.} \label{tab:ppcs}
	\setlength{\tabcolsep}{0.3mm}{
		\begin{tabular}{|c|c|c|c|c|}
			\hline
			&& \multicolumn{3}{c|}{axis range} \\
			\cline{3-5}
			name & \#points & [$x_{\min}$, $x_{\max}$]  & [$y_{\min}$, $y_{\max}$] & [$z_{\min}$, $z_{\max}$] \\ \hline
			RedandBlack & 729133 & [182,575] & [10,987] &[121,353]\\
			\hline
			Loot & 784142 & [28,380] & [7,999] &[119,473]\\
			\hline
			Solider & 1059810 & [29,389] & [7,1023]&[31,436] \\
			\hline
			LongDress & 806806 & [151,397] & [5,1012]& [87,523]\\
			\hline			
			Hhi & 900153 & [0,61875] & [0,64057]& [0,17135]\\
			\hline	
	\end{tabular}}
\end{table}

We extend \eqref{feature_pooling} to represent the local graph quality by aggregating it across all color components. Note that we can represent the point cloud in different color spaces, such as the RGB, YUV, Gaussian Color Model (GCM)~\cite{geusebroek2000color,geusebroek2001color}, etc. Referring to the overall PSNR calculation of YUV content~\cite{torlig2018novel}, it suggests the color channel weighting factors as Y:U:V = 6:1:1, assuming more sensitive perception of the luminance components.
We apply  this widely-adopted factors for pooling, i.e.,
\begin{equation}\label{gcm_pooling}
S_{{\vec{s}_k}}= \frac{1}{\gamma}\sum\nolimits_{C}\gamma_C\cdot |S_{{\vec{s}_k},C}|,
\end{equation} where $\gamma_C$ is the pooling factor of $C$-channel reflecting the importance of individual color channel during visual perception, $\gamma=\sum\nolimits_{C}\gamma_C$. We will first demonstrate the efficiency of {\sf GraphSIM} in GCM space in Sec.~\ref{sec:exp}, followed by further discussions on color spaces in Sec.~\ref{sec:ablation}.
And we also quantitatively analyze the influence of different pooling methodologies for deriving \eqref{feature_pooling} and \eqref{gcm_pooling}.

In the end, we can have the overall point cloud quality by averaging all local graph similarities:
\begin{equation}\label{overall_pooling}
\mathrm{Q}=\frac{1}{\beta}\sum\nolimits_{\vec{s}_k\in \mathbf{P}_s} S_{\vec{s}_k},
\end{equation} with $\beta$ as the total number of keypoints (and corresponding constructed local graphs) defined previously.

\section{Experimental Evaluations}\label{sec:exp}
This section evaluates the {\sf GraphSIM} and other five state-of-the-art metrics for point cloud quality prediction, using two publicly accessible point cloud database, SJTU-PCQA database in \cite{SJTUdatabase} and IRPC database in \cite{javaheri2019point}.

\subsection{Subjective Point Cloud Assessment Database}


\subsubsection{SJTU-PCQA Database}
\begin{figure}[t]
	\centering
	\subfigure[]{\includegraphics[width=0.9\linewidth]{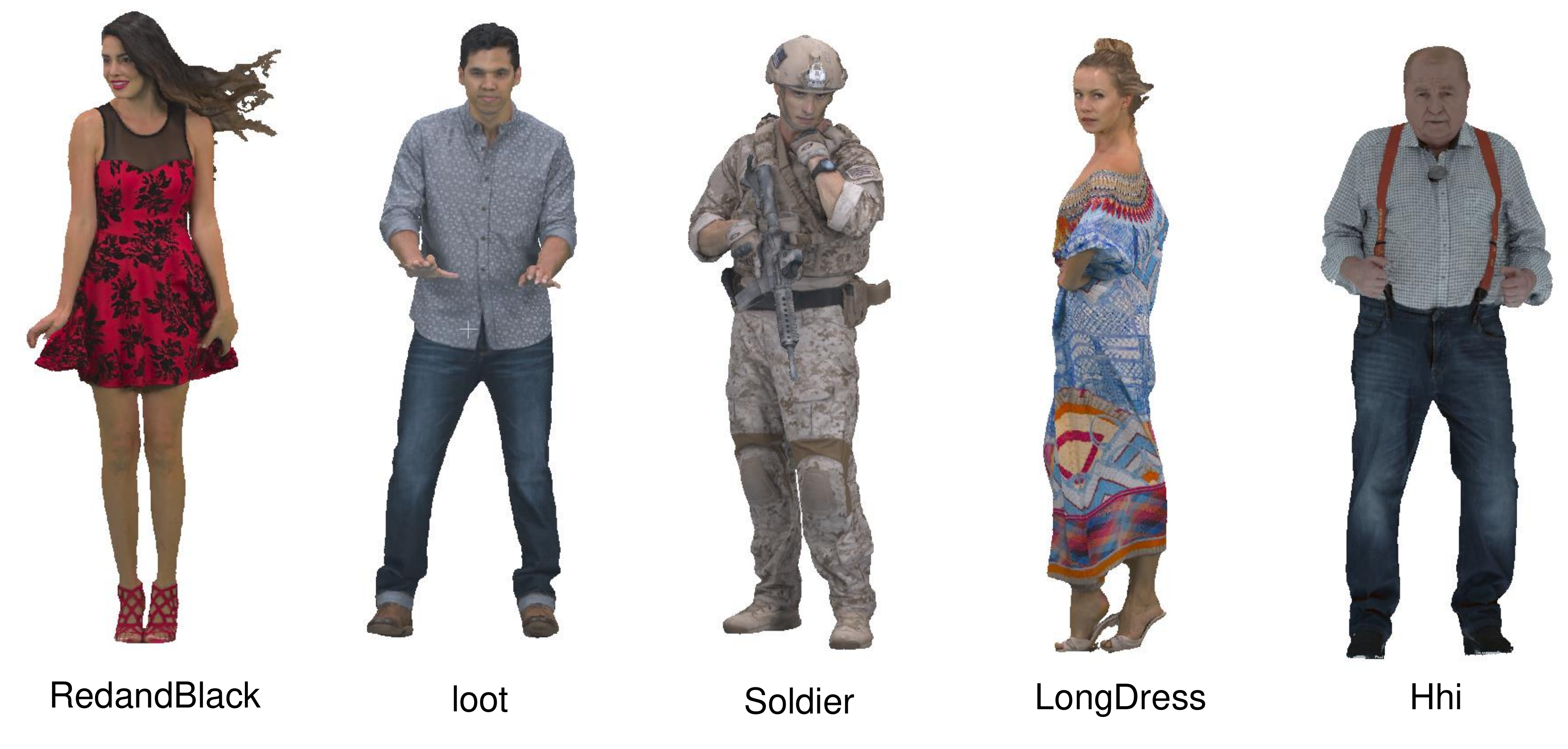} \label{database}}\\
	\subfigure[]{\includegraphics[width=0.8\linewidth]{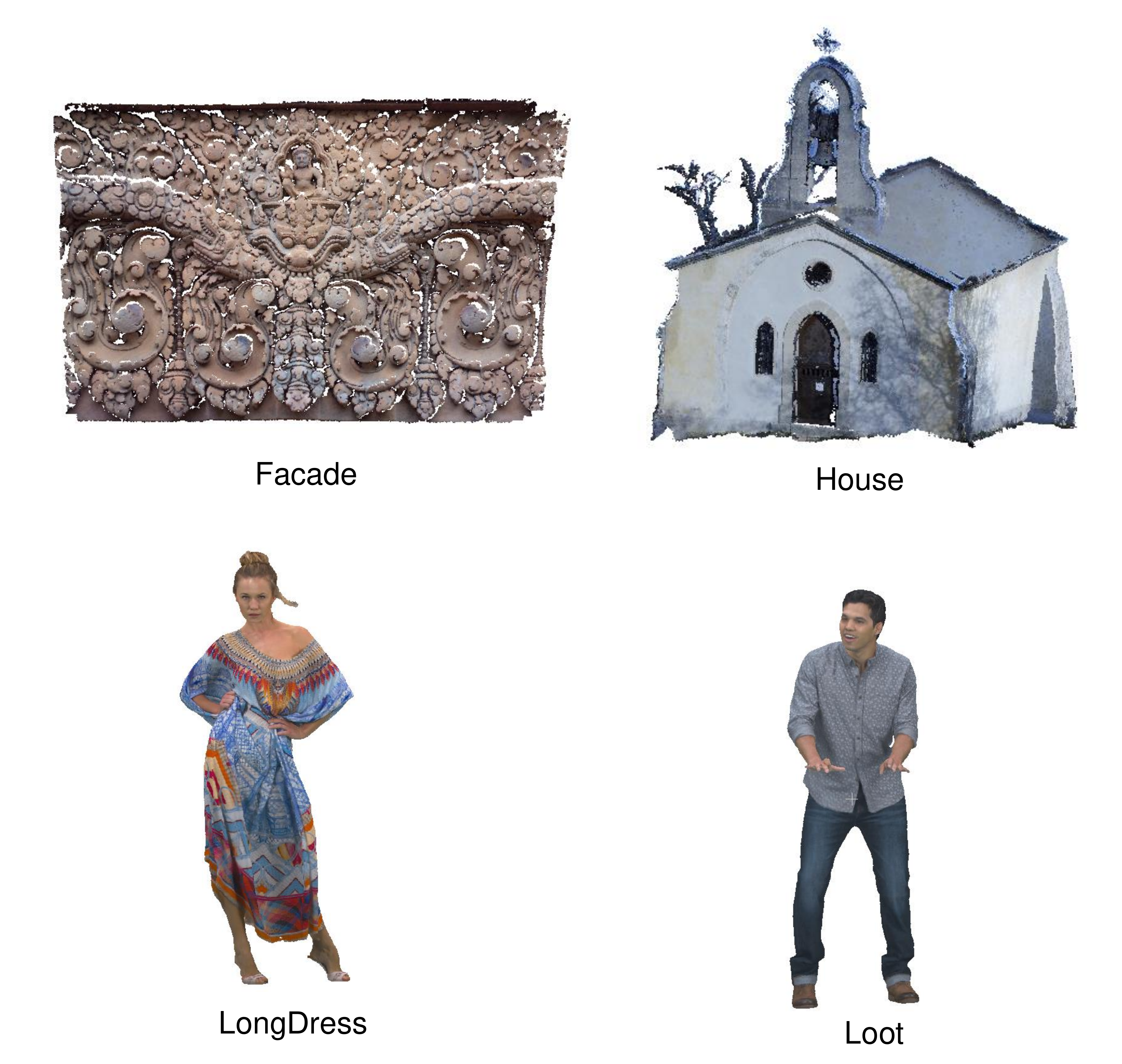} \label{irdatabase}}\\
	\caption{Point cloud databases with snapshots of original point clouds. (a) SJTU-PCQA~\cite{SJTUdatabase}; (b)  IRPC~\cite{javaheri2019point}.}
	
\end{figure}

We use five high-quality human body point cloud samples in {\bf People} category, e.g.,  ``RedandBlack'', ``Loot'', ``Soldier'', ``LongDress'', and ``Hhi''.
Table~\ref{tab:ppcs} gives the basic information of these point clouds (e.g., number of points, dimensional ranges of $x$, $y$ and $z$ axises), as well as the illustrative snapshots in Fig.~\ref{database}. These original point clouds are recommended by the experts in MPEG for compression standardization, well covering a variety of content characteristics~\cite{schwarz2018emerging}.

Each native point cloud sample is augmented with seven different types of impairments under six levels, including four individual distortions, Octree-based compression (OT), Color noise (CN), Geometry Gaussian noise (GGN), Downsampling (DS), and three superimposed distortions, such as Downsampling and Color noise (D+C), Downsampling and Geometry Gaussian noise (D+G), Color noise and Geometry Gaussian noise (C+G). These impairments, covering the re-sampling, intensity and geometric noise, and compression, are used to well simulate the artifacts that might be induced in practical applications. Please refer to~\cite{SJTUdatabase} for more details.
\subsubsection{IRPC Database}
We further adopt two high-quality {\bf Inanimate} samples, e.g., ``Facade'', ``House'', and two high-quality {\bf People} samples, ``LongDress'', ``Loot'', used in IRPC for additional evaluation shown in Fig. \ref{irdatabase}. This dataset is independently collected~\cite{javaheri2019point} with the emphasis on the point cloud compression distortions.

Each native sample is augmented using three different compression methods (e.g., OT, G-PCC and V-PCC)  with three compression levels (e.g., High quality - HQ, Medium quality - MQ, and Low quality - LQ)\footnote{Correspondingly, compression relevant quantization parameters  are respectively set at low, medium and high levels.}. More details can be found in~\cite{javaheri2019point}. We use G-PCC and V-PCC coded samples for evaluation since OT-based compression is already included in SJTU-PCQA dataset.

Interestingly, both G-PCC and V-PCC increase the points greatly after reconstruction, leading to the point density growth in a local graph.
In  Table \ref{tab:appendix}, we have noticed that it even can double the original points, e.g., G-PCC compressed ``Facade'' at HQ level. For V-PCC, additional 40\%~50\% points are reported. To the best of our knowledge, such phenomenon is barely observed for 2D image/video.

\begin{table}[t]
	\centering
	\caption{Sample Point Clouds of IPRC Database.  {\bf Inanimate} content, e.g., ``Facade'' and ``House'', is with both 12-bit and 10-bit versions; while {\bf People} content only comes with the 10-bit version. V-PCC is used to encode 10-bit samples for both {\bf Inanimate} and {\bf People} categories; G-PCC encodes 12-bit {\bf Inanimate} content, and 10-bit {\bf People} sequences as detailed in~\cite{javaheri2019point}.} \label{tab:appendix}
	\setlength{\tabcolsep}{0.2mm}{
		\begin{tabular}{|c|c|c|c|c|c|c|c|c|c|}
			\hline
			Name &\multicolumn{2}{|c|}{Type}&\#Points& $\Delta$ \% & Name &\multicolumn{2}{|c|}{Type}&\#Points& $\Delta$ \% \\ \hline
			\multirow{8}{*}{\rotatebox{90}{Facade}}&\multirow{2}{*}{Original} & 12  bit&{\bf 1596085}&-& \multirow{8}{*}{\rotatebox{90}{House}}&\multirow{2}{*}{Original} & 12 bit&{\bf4848745}&- \\ \cline{3-5} \cline{8-10}
			& & 10 bit&{\bf 889698}&-& & & 10 bit& {\bf1724175}&- \\ \cline{2-5} \cline{7-10}
			&\multirow{3}{*}{G-PCC}&HQ & 4699133&+194\%& &\multirow{3}{*}{G-PCC}&HQ & 8962169&+85\% \\ \cline{3-5} \cline{8-10}
            &&MQ & 3987755&+150\%& &&MQ& 7568556&+56\%\\ \cline{3-5} \cline{8-10}
            &&LQ & 3335666&+109\%& &&LQ& 6322527&+30\%\\ \cline{2-5} \cline{7-10}
            &\multirow{3}{*}{V-PCC}&HQ & 1246656&+40\%& &\multirow{3}{*}{V-PCC}&HQ & 2638304&+42\%\\ \cline{3-5} \cline{8-10}
            &&MQ &1246656&+40\% & &&MQ & 2638304&+42\%\\ \cline{3-5} \cline{8-10}
            &&LQ & 1246656&+40\%&&&LQ & 2638304&+42\% \\ \hline		
            \multirow{7}{*}{\rotatebox{90}{Longdress}}&{Original}&10 bit & {\bf857966}&-& \multirow{7}{*}{\rotatebox{90}{Loot}}&{Original} & 10 bit& {\bf805285}&- \\ \cline{2-5} \cline{7-10}
			&\multirow{3}{*}{G-PCC}&HQ & 1166444&+36\%& &\multirow{3}{*}{G-PCC}&HQ & 1087420&+35\% \\ \cline{3-5} \cline{8-10}
            &&MQ & 985590&+15\%& &&MQ& 915023&+14\%\\ \cline{3-5} \cline{8-10}
            &&LQ & 886830&+3\%& &&LQ& 825052&+2\%\\ \cline{2-5} \cline{7-10}
            &\multirow{3}{*}{V-PCC}&HQ & 1271456&+48\%& &\multirow{3}{*}{V-PCC}&HQ & 1217856&+51\%\\ \cline{3-5} \cline{8-10}
            &&MQ & 1271456&+48\%& &&MQ & 1217856&+51\%\\ \cline{3-5} \cline{8-10}
            &&LQ & 1271456&+48\%&&&LQ & 1217856&+51\%\\ \hline
	\end{tabular}}
\end{table}

\subsection{Gaussian Color Decomposition.}We first utilize the Gaussian Color Model (GCM) to demonstrate the model efficiency. This is mainly because GCM is suggested to be more closely related to the color sensation of our HVS~\cite{geusebroek2001color}. Other color spaces are discussed in subsequent ablation studies. Normally, the GCM function decomposes the native RGB signal via
\begin{equation}\label{transfin}\left[
\begin{array}{c}
\widehat{E} \\
\widehat{E}_{\lambda} \\
\widehat{E}_{\lambda\lambda}
\end{array}\right]=\left(
\begin{array}{ccc}
0.06 & 0.63 & 0.27 \\
0.30 & 0.04 & -0.35 \\
0.34 & -0.6 & 0.17
\end{array}\right)\cdot\left[
\begin{array}{c}
R\\
G\\
B
\end{array}\right],
\end{equation}
where $\widehat{E}$, $\widehat{E}_{\lambda}$ and $\widehat{E}_{\lambda\lambda}$ are respective luminance and two chrominance components.

\begin{table*}[pt]
	\centering
	\caption{Model performance (PLCC, SROCC and RMSE) for different point clouds categories in SJTU-PCQA database. R\&B represents ``RedandBlack'', L\&D represents ``LongDress''.} \label{Table:overall}
	\begin{scriptsize}
	\setlength{\tabcolsep}{0.7mm}{
		\begin{tabular}{|c|c|c|c|c|c|c|c|c|c|c|c|c|c|c|c|c|c|c|c|}
			\hline
			\multicolumn{2}{|c|}{} & \multicolumn{6}{|c|}{PLCC} & \multicolumn{6}{|c|}{SROCC} & \multicolumn{6}{|c|}{RMSE}  \\ \hline
			\multicolumn{2}{|c|}{metric:}&R\&B&Loot&Soldier&L\&D&Hhi&{\bf ALL}&R\&B&Loot&Soldier&L\&D&Hhi&{\bf ALL}&R\&B&Loot&Soldier&L\&D&Hhi&{\bf ALL} \\  \hline
			 \multirow{2}{*}{M}& p2po &{\color{red}\textbf{0.91}}&{\color{red}\textbf{0.87}}&{\color{red}\textbf{0.92}}&0.90&0.88&{\color{red}\textbf{0.89}}&0.83&0.77&0.80&0.79&0.79&0.79&{\color{red}\textbf{0.93}}&{\color{red}\textbf{1.19}}&{\color{red}\textbf{0.99}}&1.05&1.24&{\color{red}\textbf{1.11}}\\  \cline{2-20}
			  & p2pl &0.80&0.75&0.80&0.81&0.54&0.74&0.75&0.69&0.71&0.73&0.56&0.66&1.38&1.62&1.48&1.43&2.23&1.66 \\ \cline{1-20}
			 \multirow{2}{*}{H} & p2po &0.83&0.69&0.80&0.73&0.79&0.80&0.74&0.67&0.70&0.72&0.72&0.70&1.29&1.86&1.51&1.76&1.62&1.49  \\ \cline{2-20}
			  & p2pl &0.80&0.70&0.71&0.80&0.75&0.71&0.74&0.60&0.63&0.72&0.67&0.66&1.38&1.74&1.84&1.46&1.76&1.83 \\ \cline{1-20}
			 \multicolumn{2}{|c|}{$\mathrm{PSNR_{YUV}}$}   &0.70&0.73&0.72&0.90&0.79&0.71&0.71&0.71&0.71&0.90&0.73&0.71&1.64&1.67&1.72&1.04&1.66&1.74 \\ \cline{1-20}
			 \multicolumn{2}{|c|}{\textbf{GraphSIM}}   &0.86&0.86&0.91&{\color{red}\textbf{0.95}}&{\color{red}\textbf{0.90}}&{\color{red}\textbf{0.89}}&{\color{red}\textbf{0.86}}&{\color{red}\textbf{0.87}}&{\color{red}\textbf{0.89}}&{\color{red}\textbf{0.94}}&{\color{red}\textbf{0.88}}&{\color{red}\textbf{0.88}}&1.17&1.24&1.02&{\color{red}\textbf{0.73}}&{\color{red}\textbf{1.17}}&1.13  \\ \hline
	\end{tabular}}
	\end{scriptsize}
\end{table*}

\begin{table*}[pt]
	\centering
	\caption{Model performance (PLCC, SROCC and RMSE) for point clouds samples in SJTU-PCQA database in terms of different impairments. {\bf People(ave)} represents treat all the samples in SJTU-PCQA as whole.} \label{Table:single}
	\begin{scriptsize}
	\setlength{\tabcolsep}{0.7mm}{
		\begin{tabular}{|c|c|c|c|c|c|c|c|c|c|c|c|c|c|c|c|c|c|c|c|c|c|c|c|c|c|c|}
			\hline
			\multicolumn{3}{|c|}{} & \multicolumn{8}{|c|}{PLCC} & \multicolumn{8}{|c|}{SROCC} & \multicolumn{8}{|c|}{RMSE}  \\ \hline
			\multicolumn{3}{|c|}{metric:}&OT&CN&GGN&DS&D+C&D+G&C+G&{\bf ALL} & OT&CN&GGN&DS&D+C&D+G&C+G&{\bf ALL} & OT&CN&GGN&DS&D+C&D+G&C+G&{\bf ALL} \\  \hline
			\multirow{6}{*}{\rotatebox{90}{People (ave)}} & \multirow{2}{*}{M}  & p2po &0.88&-&0.97&0.96&0.97&0.99&0.99&{\color{red}\textbf{0.89}}&0.80&-&0.95&0.92&0.97&0.97&0.97&0.79&0.84&-&0.67&0.64&0.61&0.38&0.40&{\color{red}\textbf{1.11}} \\ \cline{3-27}
			&  & p2pl &0.90&-&0.96&0.80&0.77&0.99&0.99&0.74&0.80&-&0.94&0.56&0.66&0.98&0.98&0.66&0.78&-&0.71&1.36&1.55&0.34&0.39&1.66  \\ \cline{2-27}
			& \multirow{2}{*}{H} & p2po &0.86&-&0.96&0.91&0.84&0.99&0.99&0.80&0.80&-&0.95&0.89&0.81&0.97&0.97&0.70&0.92&-&0.66&0.95&1.32&0.38&0.40&1.49  \\ \cline{3-27}
			&  & p2pl &0.89&-&0.97&0.82&0.86&0.99&0.99&0.71&0.81&-&0.95&0.77&0.87&0.97&0.97&0.66&0.82&-&0.67&1.31&1.23&0.37&0.40&1.83  \\ \cline{2-27}
			& \multicolumn{2}{|c|}{$\mathrm{PSNR_{YUV}}$}   &0.54&0.97&0.86&0.74&0.98&0.85
			&0.98&0.71&0.52&0.94&0.82
			&0.74&0.97&0.77&0.97
			&0.71&1.52&0.48&1.34&1.56&0.49
			&1.37&0.52&1.74 \\ \cline{2-27}
			& \multicolumn{2}{|c|}{\textbf{GraphSIM}}   &0.81&0.90&
			0.97&0.97&0.95&0.99&0.98&{\color{red}\textbf{0.89}}&0.71&0.82&
			0.96&0.91&0.95&0.96&0.97&{\color{red}\textbf{0.88}}&1.05&0.78&0.62&0.55
			&0.79&0.43&0.52&1.13  \\ \hline
	\end{tabular}}
	\end{scriptsize}
\end{table*}

\subsection{Model Parameters.} In total, we have a set of parameters associated with different processing stages that need to be determined for {\sf GraphSIM}.
\begin{itemize}
	\item $\beta$, $L$ in {\it Resampling}: $\beta=N/1000$ and  $L$ = 4. These numbers are used to well balance the efficiency and complexity.
	\item $\theta$, $\tau$, $\sigma$ in {\it Local Graph Construction}: Given a local graph centered at $\vec{s}_k$, we set $\theta=\frac{1}{10}\mathrm{B}$ to cluster neighbors, where $\mathrm{B}=\min$($X_s$,$Y_s$,$Z_s$) with $X_s={X_{\max}-X_{\min}}$, $Y_s=Y_{\max}-Y_{\min}$ and $Z_s=Z_{\max}-Z_{\min}$ as respective bounding box scale of $x$-, $y-$, and $z$-axis of reference point; We then determine the $\tau$ using the largest Euclidean distance from the 50 nearest neighbors of all the points in $\vec{\bf V}_{(\vec{\bf X}_r, \vec{s}_k)}$ or $\vec{\bf V}_{(\vec{\bf X}_d, \vec{s}_k)}$ in \eqref{local set}. If there are less than 50 neighbors (e.g., $<$ 50 points in $\vec{\bf V}_{(\vec{\bf X}_r, \vec{s}_k)}$ or $\vec{\bf V}_{(\vec{\bf X}_d, \vec{s}_k)}$), the largest Euclidean distance is set as $\tau$; Finally, we have $\sigma^2=\tau^2/2$. It is worth to point out that we actually connect the $\tau$ and $\theta$ of local graph with the scale of point cloud (e.g., bounding box size). This allows us to perform the  normalization evenly (without bias) to the same scale for fair comparison.
	\item $T_0$, $T_1$, $T_2$, $\gamma_C$ in {\it Similarity Pooling}: $T_0$, $T_1$, and $T_2$ are set as 0.001 following the suggestion in~\cite{wang2004ssim};  We set $[\gamma_{\widehat{E}},\gamma_{\widehat{E}_{\lambda}}, \gamma_{\widehat{E}_{\lambda\lambda}}] = [6,1,1]$ to reflect the different importance of various color components where normally, luminance is more sensitive to the HVS. The numbers, e.g., 6, 1, and 1, follow the conventions widely used in compression society to derive the overall PNSR of all YUV channels where weighting coefficients for Y, U, and V are 6, 1, and 1, i.e., $[\gamma_{\rm Y},\gamma_{\rm U},\gamma_{\rm V}]=[6,1,1]$.
\end{itemize}

These parameters are either fixed constants, or can be easily derived according to the signal statistics (e.g., point cloud bounding box scale, color space, sampling ratio), suggesting that our {\sf GraphSIM} is fairly lightweight and straightforward for practical applications.

\subsection{Performance Evaluation.} We compare our {\sf GraphSIM} with another five state-of-the-art metrics adopted in MPEG point cloud compression software \cite{MPEGSoft}, e.g.,
\begin{itemize}
    \item PSNR-MSE-P2point (M-p2po)
    \item PSNR-MSE-P2plane (M-p2pl)
    \item PSNR-Hausdorff-P2point (H-p2po)
    \item PSNR-Hausdorff-P2plane (H-p2pl)
    \item $\mathrm{PSNR_{YUV}}$: $\mathrm{PSNR_{YUV}}=(6\times \mathrm{PSNR_Y}+\mathrm{PSNR_U}+\mathrm{PSNR_V})/8$) \cite{torlig2018novel}
\end{itemize}
Note that first four metrics only give the geometry measurement without color components using either point-to-point (p2po) or point-to-plane (p2pl) error-based Hausdorff or MSE (mean squared error) distances, while $\mathrm{PSNR_{YUV}}$ calculates the overall YUV distortion of matched points in reference and distorted contents.

To ensure the consistency between subjective scores (e.g., MOS) and objective predictions from various  models, we map the objective predictions of different models to the same dynamic range following the recommendations suggested by the  video quality experts group (VQEG) \cite{video2003final,sheikh2006statistical}, to derive popular PLCC for prediction accuracy, SROCC for prediction monotonicity, and RMSE for prediction consistency for evaluating the model performance. The larger PLCC or SROCC comes with the better model performance. On the contrary, the lower RMSE is better.
More details can be found in~\cite{video2003final}.
\begin{figure}[pt]
	\centering
	\subfigure[]{		\includegraphics[width=0.48\linewidth]{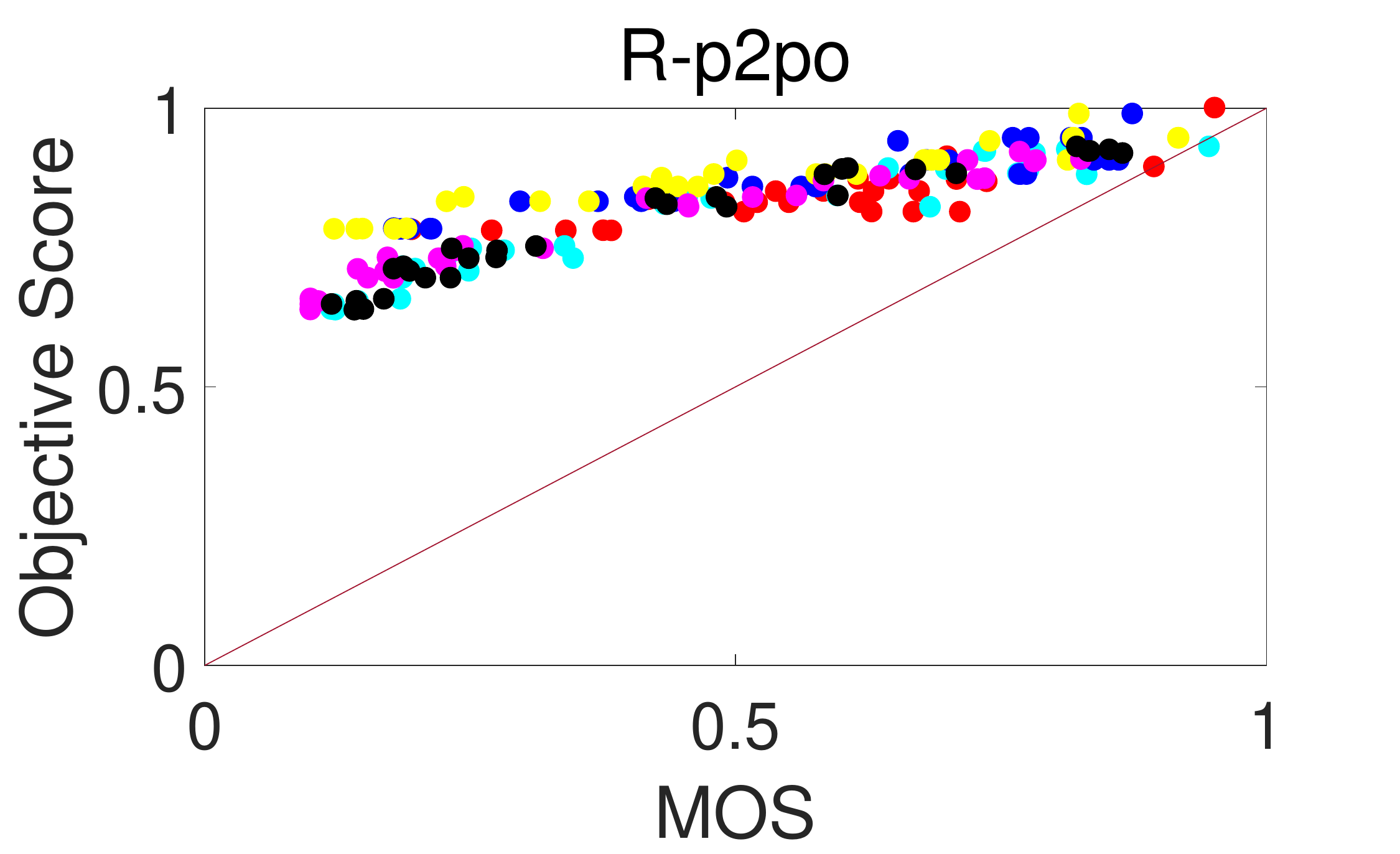}		\label{a}	}%
	\subfigure[]{		\includegraphics[width=0.48\linewidth]{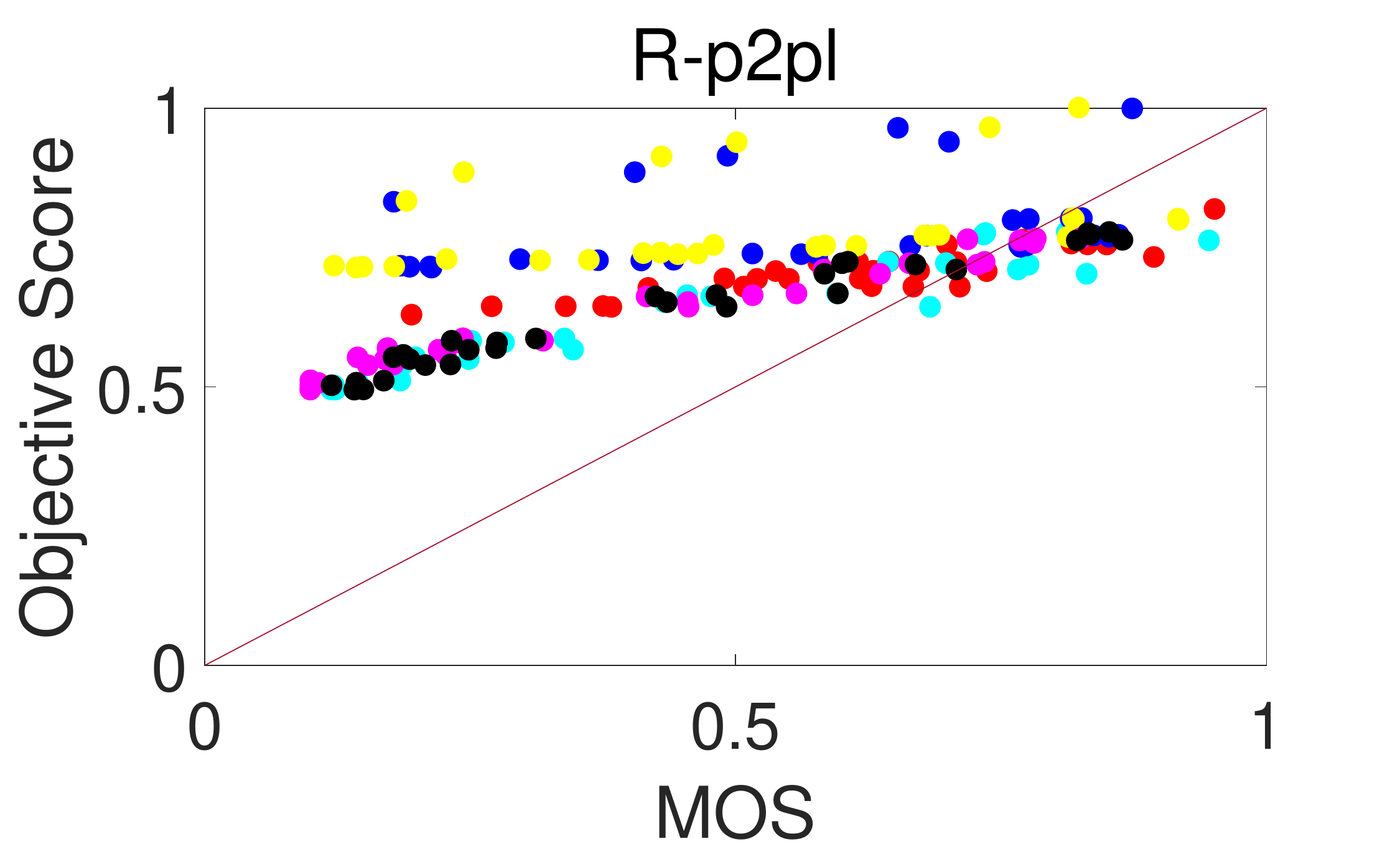}		\label{b}	}\\
	\subfigure[]{		\includegraphics[width=0.48\linewidth]{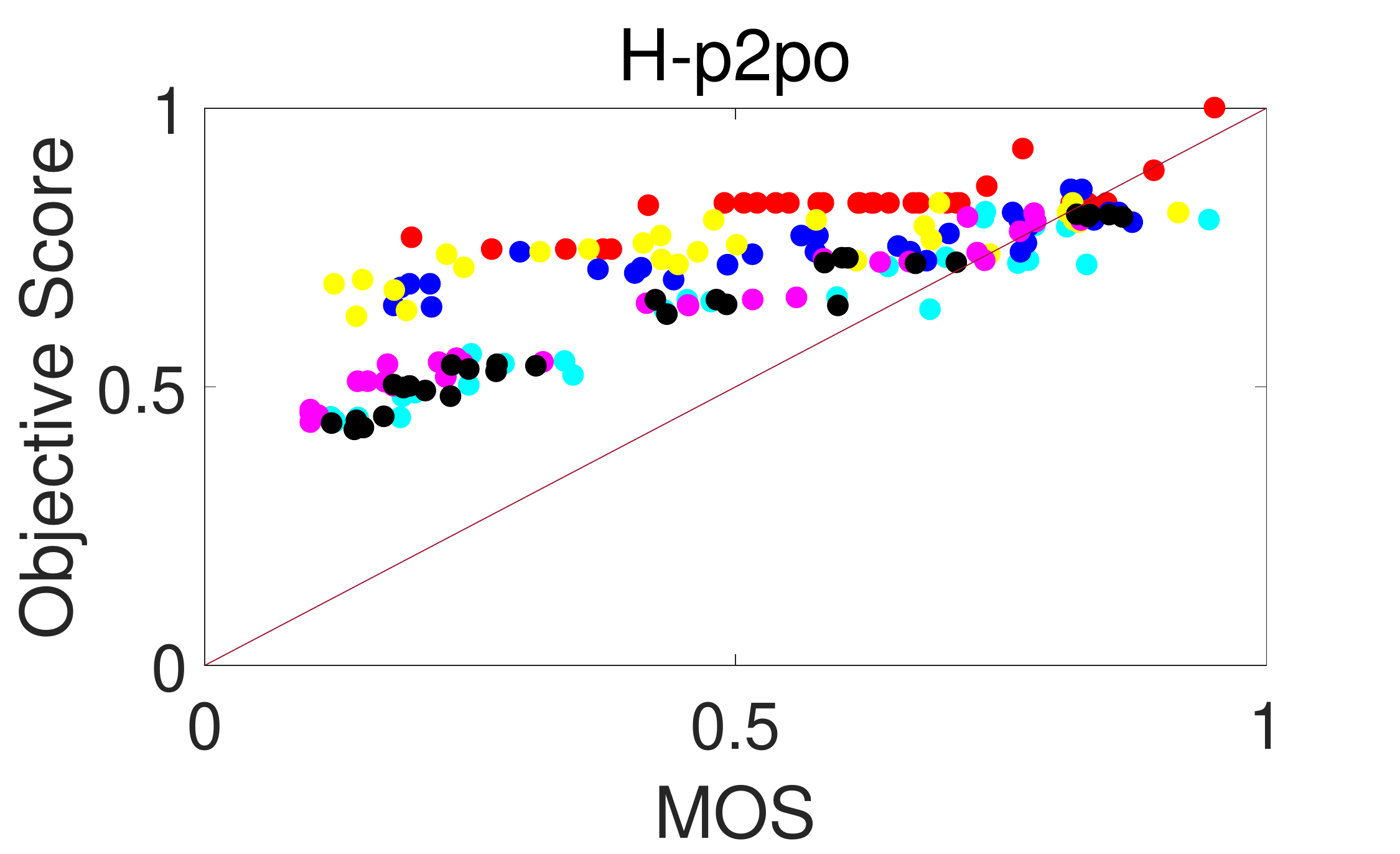}		\label{c}	}%
	\subfigure[]{		\includegraphics[width=0.48\linewidth]{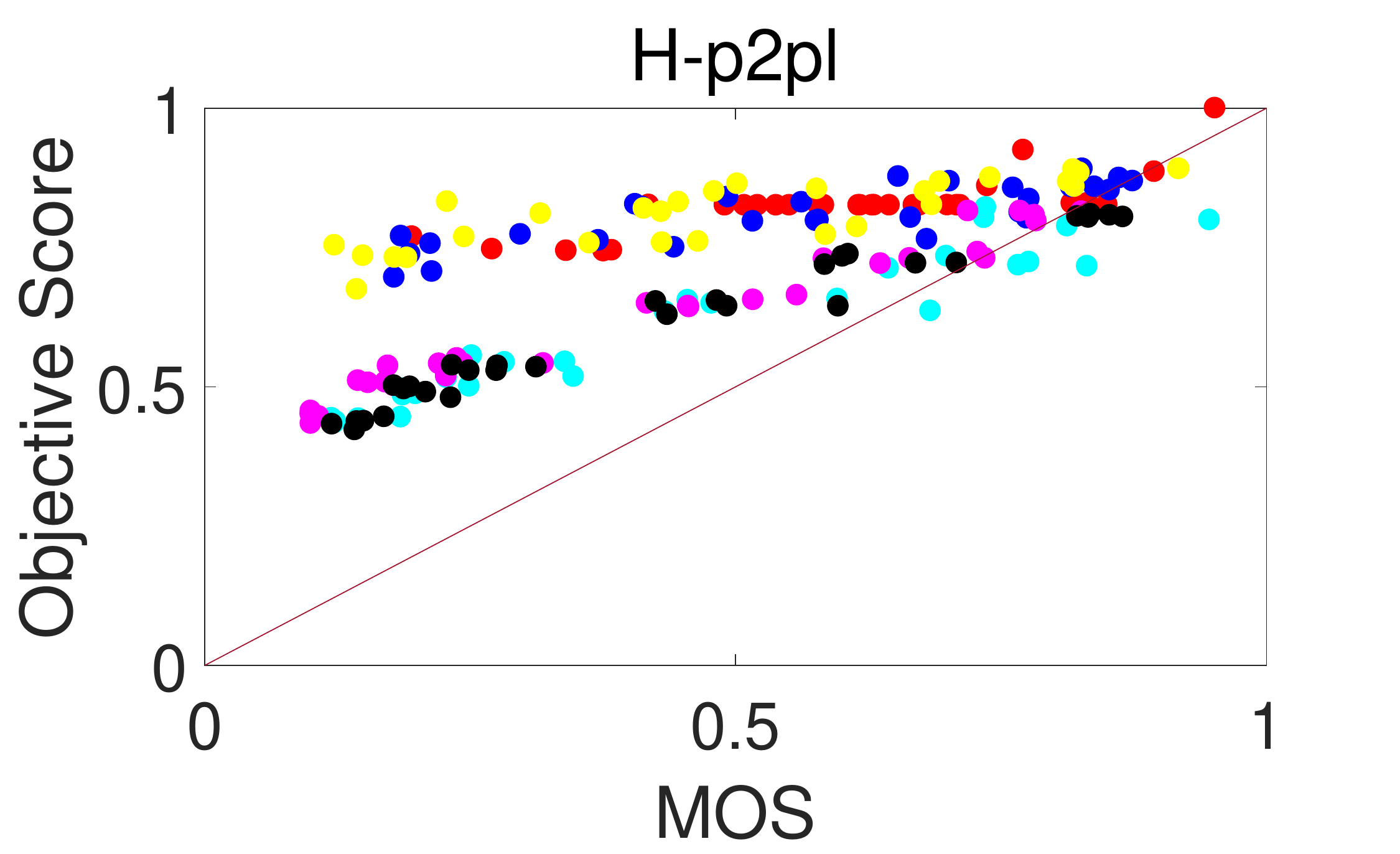}		\label{d}	} \\
	\subfigure[]{		\includegraphics[width=0.48\linewidth]{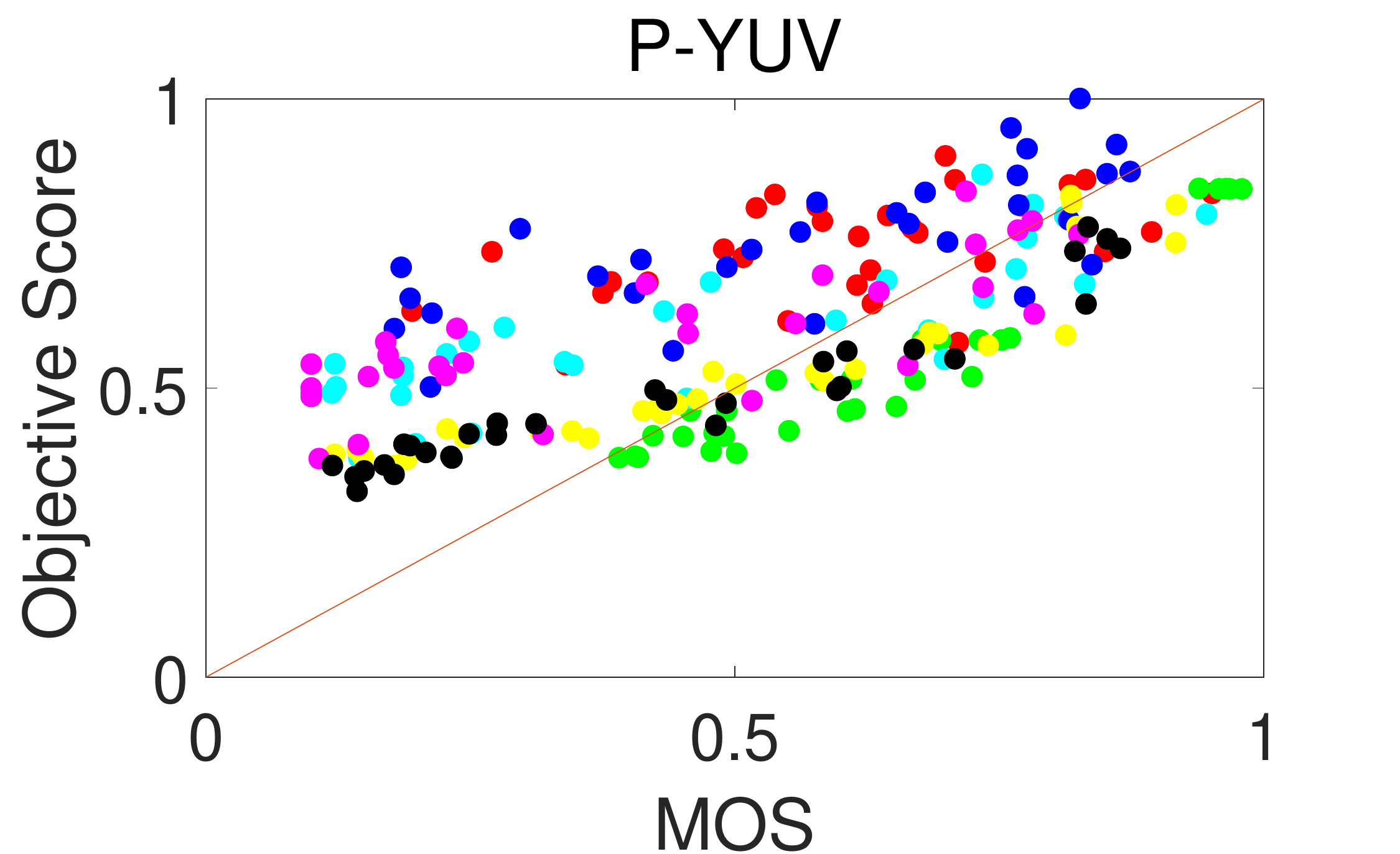}		\label{e}	}%
	\subfigure[]{		\includegraphics[width=0.48\linewidth]{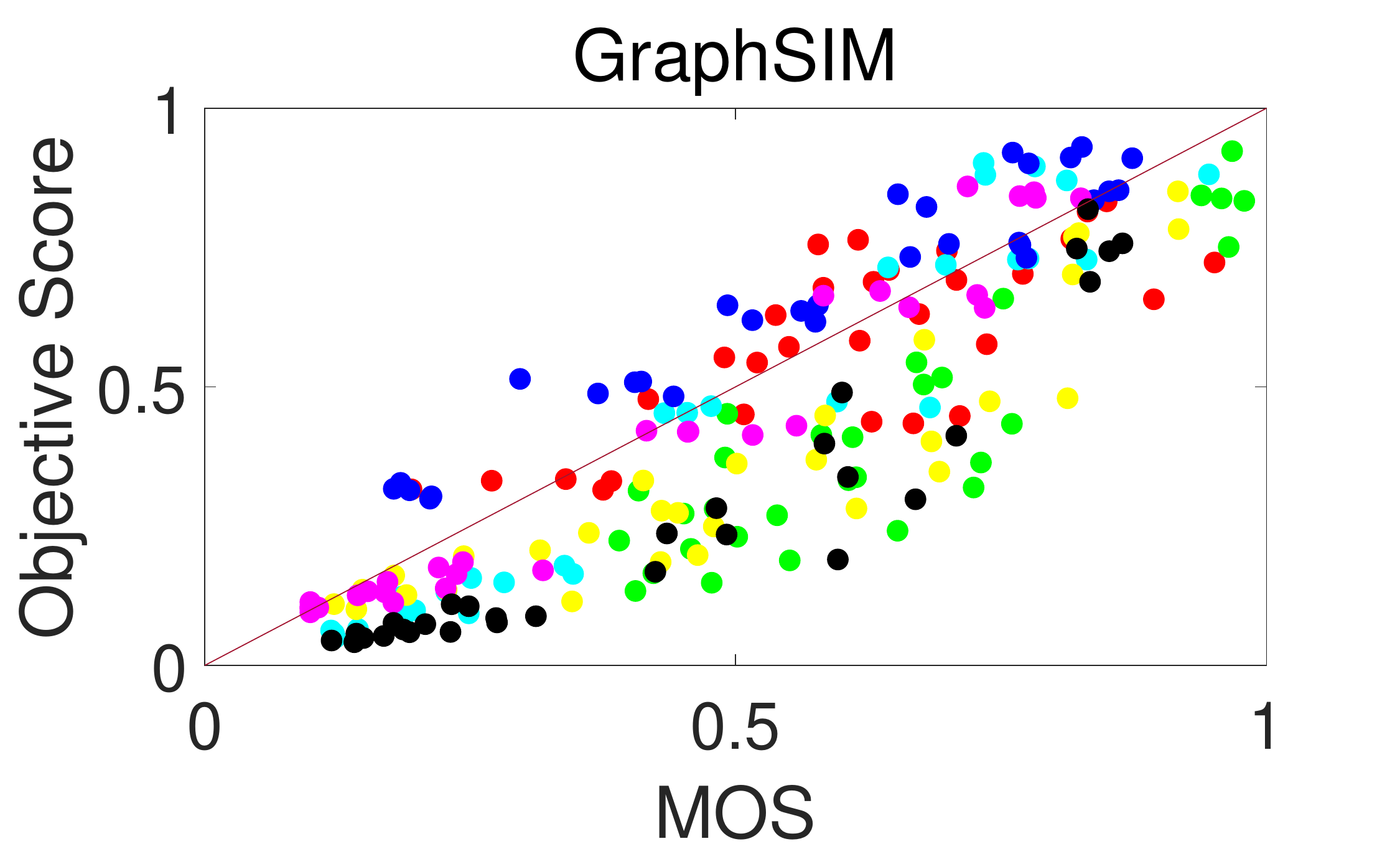}		\label{f}	}\\
	\subfigure{		\includegraphics[width=0.60\linewidth]{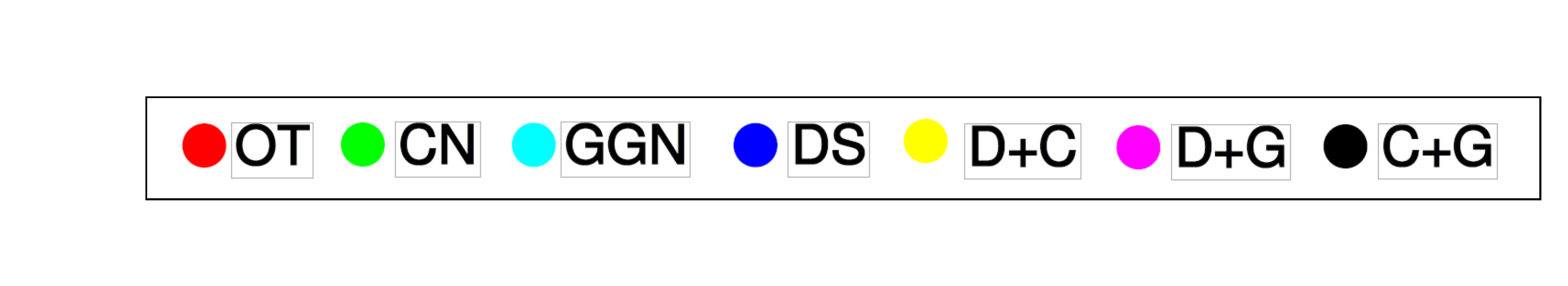}	}%
	\caption{MOS prediction accuracy of objective metrics: subplots (a)-(d) are point-wise distance-based metrics used for geometry distortion without color attributes (no CN points); subplots (e)(f) consider color attributes (with CN points); solid line is ``$y=x$'' implying the perfect prediction when being overlapped with this line.}
	\label{fig:scatter_plots}
\end{figure}
\subsubsection{\bf SJTU-PCQA Database}
In this part, we present the performance of {\sf GraphSIM} and other state-of-the-art metrics over SJTU-PCQA database.
As reported in Table \ref{Table:overall}, our {\sf GraphSIM} consistently offers the leading performance (mostly ranked at the top), in {\bf People} categories for all impairments, with model performance evaluated using (PLCC, SROCC, RMSE) as (0.89, 0.88, 1.13). In comparison, model performances are (0.89, 0.79, 1.11) for metric M-p2po, (0.74, 0.66, 1.66) for M-p2pl,  (0.80, 0.70, 1.49)  for H-p2po, (0.71, 0.66, 1.83)  for H-p2pl  and  (0.71, 0.71, 1.74) for PSNR$_{\rm YUV}$, respectively.

{\bf Consistency.} The {\sf GraphSIM} shows robust  performance across various contents and impairments, and demonstrates reliable correlations with subjective MOSs. On the contrary, other metrics varies quite significantly for different cases. For example, objective metric M-p2po presents comparable PLCC to {\sf GraphSIM} for ``Loot'' (0.87 vs 0.86), ``Soldier'' (0.92 vs 0.91) and overall {\bf People} category (0.89 vs 0.89), but exhibits obvious inferior correlation for SROCC, e.g., (0.77 vs 0.88), (0.80 vs 0.89) and (0.79 vs 0.88), respectively. One potential cause for such unreliable variation is due to the normalization scheme.  As for M-p2po metric,
though we calculate the distance between paired points in respective reference and distorted point clouds,
distance itself varies significantly for different points clouds because of the dimensional range scale differences. For example, the axis range $[(x_{\min}, x_{\max}), (y_{\min}$, $y_{\max}),  (z_{\min}$, $z_{\max})]$ of ``Hhi''
is
[(0,61875), (0,64057), (0,170135)], while the axis range of ``Loot'' is [(28,380), (7,999), (119,473)]. This would apparently lead to the variations of distance measurement in several order of magnitudes. To overcome this issue, a simple fix is proposed to  normalize the distance-based MSE using the maximum dimensional range $p$, e.g., $p=\max[(x_{\max}-x_{\min}),(y_{\max}-y_{\min}),(z_{\max}-z_{\min})]$~\cite{tian2017geometric}. It, however, still can not avoid the bias if the  distance error is no aligned with axis having the maximum range. All of these suggest that point-wise distance-based metrics are difficult to be generalized for reliable performance measurement.

{\bf Impairment Superimposition.} We present the performance of metrics in terms of different distortion in Table \ref{Table:single}. Except for the {\sf GraphSIM} and PSNR$_{\rm YUV}$, other four metrics only consider the geometrical distortion and cannot handle color attribute impairments, e.g., CN (color noise), contrast change, etc, at all.  This limits the generalization of these distance-based metrics, e.g., M-p2po, M-p2pl, H-p2po, and H-p2pl, for evaluating the superimposed distortions. The same problem is observed for PSNR$_{\rm YUV}$ as well.
The PSNR$_{\rm YUV}$ presents the best performance for CN distortion on average, having (PLCC, SROCC, RMSE) =  (0.97, 0.94, 0.48). It, however, offers the worst performance for OT impairment with (PLCC, SROCC, RMSE) = (0.54, 0.52, 1.52). This is mainly because  PSNR$_{\rm YUV}$ only calculates the color intensity difference for two matched points from both reference and impaired point clouds, without considering the geometric impacts. As quantitatively listed in Table \ref{Table:single},  PSNR$_{\rm YUV}$ offers relatively poor performance for cases with geometric distortion, such as GGN with (PLCC, SROCC, RMSE) = (0.86, 0.82, 1.34), DS with (PLCC, SROCC, RMSE) = (0.74, 0.74, 1.56), D+G with (PLCC, SROCC, RMSE) = (0.85, 0.77, 1.37)). In contrast, {\sf GraphSIM} provides the (PLCC, SROCC, RMSE) at (0.97, 0.96, 0.62), (0.97, 0.91, 0.55,) and (0.99, 0.96, 0.43) for corresponding GGN, DS, and D+G, respectively. Besides, we give some illustrative examples of distorted samples in Fig. \ref{fig:mos_example}. Corresponding distortion levels, MOSs and GraphSIM scores are provided.




{\bf Scatter Plot.}  For better illustration, we have also provided the scatter plots shown in Fig.~\ref{fig:scatter_plots} for all six models. Though some metrics provide better performance in certain types of impairments (e.g., M-p2po for ``C+G'' distortion), they are not reliable and consistent. This is also reflected from the scatter plots. All point-wise distance-based metrics could not offer competitive performance with our {\sf GraphSIM} in predicting the subjective MOS, where most predictions are away from the  ``$y=x$'' axis (e.g., perfect-prediction line). 

\subsubsection{\bf IRPC Database}
We further evaluate {\sf GraphSIM} using another IRPC database shown in Table \ref{TABLE-TTRPC}. 
Overall, the {\sf GraphSIM} provides the noticeable performance margin with (PLCC, SROCC, RMSE) = (0.94, 0.76, 0.21)  for all samples in both {\bf Inanimate} and {\bf People} categories, as shown in Table~\ref{TABLE-TTRPC}.

Performance consistency is still  a critical issue for other metrics. For example, though PSNR$_{\rm YUV}$ shows the similar PLCC and RMSE as the GraphSIM, it has severely degradation in SROCC measurement even for {\bf People} contents. We then retrieve the MOS and respective objective scores in Table~\ref{TABLE-obj}.  As we can see,  $\mathrm{PSNR_{YUV}}$ exhibits  larger variations across different content. For example, $\mathrm{PSNR_{YUV}}$ of ``Loot''
at LQ scale even offers higher objective score than it of ``Longdress'' at the HQ scale.  This comes from the lack of inappropriate geometric scale normalization since $\mathrm{PSNR_{YUV}}$  only applies the error measurement between matched points. The same inconsistency is observed for $\mathrm{PSNR_{YUV}}$ in evaluating the OT artifacts shown in Table~\ref{Table:single}.

\begin{table}[pt]
	\centering
	\caption{Model performance (PLCC, SROCC and RMSE) for different point clouds encoded using G-PCC  and V-PCC.} \label{TABLE-TTRPC}
	\begin{scriptsize}
	\setlength{\tabcolsep}{0.4mm}{
		\begin{tabular}{|c|c|c|c|c|c|c|c|c|c|c|c|}
			\hline
			\multicolumn{3}{|c|}{} & \multicolumn{3}{|c|}{PLCC} & \multicolumn{3}{|c|}{SROCC} & \multicolumn{3}{|c|}{RMSE}  \\ \cline{1-12}
			\multicolumn{3}{|c|}{metric:}&G-PCC&V-PCC&{\bf ALL}&G-PCC&V-PCC&{\bf ALL}&G-PCC&V-PCC&{\bf ALL} \\  \hline
\multirow{6}{*}{\rotatebox{90}{People}}&\multirow{2}{*}{M}
			 &p2po &0.99&1.00&0.97&0.94&0.99&{\color{red}\textbf{0.88}}&0.06&0.03&0.10  \\ \cline{3-12}
			&  & p2pl &0.98&0.99&0.95&0.94&0.99&{\color{red}\textbf{0.88}}&0.07&0.04&0.11 \\ \cline{2-12}
			& \multirow{2}{*}{H} & p2po &0.37&0.97&0.52&0.17&0.88&0.33&0.40&0.07&0.32  \\ \cline{3-12}
			&  & p2pl &0.98&0.83&0.95&0.88&0.70&0.78&0.07&0.17&0.11\\ \cline{2-12}
			& \multicolumn{2}{|c|}{$\mathrm{PSNR_{YUV}}$}   &0.99&0.98&0.95&0.77&0.29&0.51&0.02&0.10&0.12 \\ \cline{2-12}
			& \multicolumn{2}{|c|}{\textbf{GraphSIM}}   &0.98&1.00&{\color{red}\textbf{0.98}}&0.71&0.93&0.85&
			0.08&0.03&{\color{red}\textbf{0.07}}  \\ \hline		
\multirow{6}{*}{\rotatebox{90}{Inanimate}}&\multirow{2}{*}{M}
			 &p2po &0.96&0.87&0.89&0.94&-0.49&0.48&0.23&0.09&0.29  \\ \cline{3-12}
			&  & p2pl &0.98&0.34&0.89&0.94&-0.12&{\color{red}\textbf{0.52}}&0.15&0.17&0.28 \\ \cline{2-12}
			& \multirow{2}{*}{H} & p2po &0.60&0.96&0.65&0.37&0.61&0.46&0.65&0.05&0.48  \\ \cline{3-12}
			&  & p2pl &0.95&0.78&0.62&0.43&-0.75&0.21&0.25&0.11&0.49 \\ \cline{2-12}
			& \multicolumn{2}{|c|}{$\mathrm{PSNR_{YUV}}$}   &0.65&0.97&0.54&0.77&-0.64&0.25&0.61&0.04&0.53 \\ \cline{2-12}
			& \multicolumn{2}{|c|}{\textbf{GraphSIM}}   &0.98&0.29&{\color{red}\textbf{0.91}}&0.71&-0.15&{\color{red}\textbf{0.52}}&
			0.16&0.17&{\color{red}\textbf{0.26}}  \\ \hline		
\multirow{6}{*}{\rotatebox{90}{All}}&\multirow{2}{*}{M}
			 &p2po &0.80&0.86&0.71&0.59&-0.54&0.24&0.46&0.17&0.42  \\ \cline{3-12}
			&  & p2pl &0.39&0.93&0.47&0.28&-0.46&0.11&0.71&0.13&0.54 \\ \cline{2-12}
			& \multirow{2}{*}{H} & p2po &0.69&0.87&0.62&0.61&0.31&0.42&0.56&0.17&0.48  \\ \cline{3-12}
			&  & p2pl &0.85&0.90&0.75&0.81&-0.62&0.42&0.41&0.15&0.40 \\ \cline{2-12}
			& \multicolumn{2}{|c|}{$\mathrm{PSNR_{YUV}}$}   &0.76&0.54&0.63&0.71&0.32&0.50&0.51&0.28&0.47 \\ \cline{2-12}
			& \multicolumn{2}{|c|}{\textbf{GraphSIM}}   &0.82&0.91&{\color{red}\textbf{0.94}}&0.83&0.79&{\color{red}\textbf{0.76}}&
			0.46&0.14&{\color{red}\textbf{0.21}}  \\ \hline			
	\end{tabular}}
	\end{scriptsize}
\end{table}




\begin{table}[pt]
	\centering
	\caption{Objective score (PSNR$_{\rm YUV}$, {\sf GraphSIM}) and MOS of {\bf People} samples in IRPC~\cite{javaheri2019point}.} \label{TABLE-obj}
	\setlength{\tabcolsep}{0.7mm}{
		\begin{tabular}{|c|c|c|c|c|c|c|c|}
			\hline
			\multicolumn{2}{|c|}{} & \multicolumn{3}{|c|}{LongDress} & \multicolumn{3}{|c|}{Loot}   \\ \hline
			\multicolumn{2}{|c|}{level:}&HQ&MQ&LQ&HQ&MQ&LQ \\  \hline
			 \multirow{2}{*}{$\mathrm{PSNR_{YUV}}$}
			 &G-PCC &38.7&39.0&32.2&49.3&48.3&41.5  \\ \cline{2-8}
			  & V-PCC &40.7&39.2&35.8&51.1&49.0&45.4 \\ \cline{1-8}
\multirow{2}{*}{{\sf GraphSIM}}
			 &G-PCC &0.86&0.83&0.70&0.85&0.82&0.73  \\ \cline{2-8}
			  & V-PCC &0.86&0.83&0.75&0.86&0.82&0.74\\ \cline{1-8}
			 \multirow{2}{*}{MOS} &G-PCC  &4.50&4.45&3.65&4.70&4.60&3.70  \\ \cline{2-8}
			  & V-PCC &4.65&4.50&3.90&4.55&4.40&3.90 \\ \hline			
	\end{tabular}}
\end{table}



\section{Ablation Studies} \label{sec:ablation}
This section have examined the {\sf GraphSIM} by dissecting and reassembling its modules to demonstrate its generalization and efficiency.

\subsection{Color Space}

In Sec.~\ref{sec:exp}, we have exemplified the efficiency of {\sf GraphSIM} assuming the GCM-based color channel decomposition. It  mainly follows the suggestions that GCM well correlates with the color sensation of our HVS~\cite{geusebroek2001color}. In practices, we may use other color spaces, such as RGB and YUV that are typically applied in compression societies. We set the same color weighting factors for YUV and GCM spaces, e.g., $\gamma_Y$ = 6, $\gamma_U$ = 1 and $\gamma_V$=1, given that luminance component is more sensitive~\cite{torlig2018novel};   For RGB space,  $\gamma_R$ = 1, $\gamma_G$ = 2 and $\gamma_V$= 1. It follows the observations that green color components are more sensible to our vision system. The exact weighting factor setting of RGB is motivated by the fact that, in typical imaging CMOS, we often have two green pixels associated one red and one blue pixel.  Table~\ref{tab:color_space} lists the model performance averaged for all sequences in {\bf People} category across GCM, RGB, and YUV. Other modules in {\sf GraphSIM} are kept as suggested in Sec.~\ref{sec:exp}. As reported, our {\sf GraphSIM} has shown consistent performance across various color spaces, again ensuring the model generalization to different applications.

\begin{table}[t]
	\centering
	\caption{Model performance with various color spaces.} \label{tab:color_space}
	\begin{tabular}{|c|c|c|c|}
		\hline
		Color Space & PLCC & SROCC & RMSE \\ \cline{1-4}
		RGB & 0.8948 & 0.8899 & 1.1020 \\ \cline{1-4}
		YUV & 0.8941 & 0.8870 & 1.1057 \\ \cline{1-4}
		GCM & 0.8896 & 0.8842 & 1.1274 \\ \cline{1-4}
	\end{tabular}
\end{table}

\subsection{Local Graph}

{\bf Resampling.} Keypoints resampling play a vital role in {\sf GraphSIM} for maintaining the 3D geometric structure used in subsequent graph similarity measurement. We have exemplified graph filtering-based high spatial-frequency (Haar-alike high-pass filter) resampling~\cite{chen2017fast} previously. Here we introduce an additional {\it random resampling} for comparative study. In the meantime, we provide more simulations with respect to different resampling ratios (e.g., $\beta$s) for two methods as well. Table~\ref{tab:resampling_ablation} has shown the reliable performance (e.g., with outstanding PLCC, SROCC, and RMSE reported) of {\sf GraphSIM} for both methods at various sampling ratios.
\begin{table}[]
	\centering
	\caption{Model performance with different resampling mechanism: SJTU-PCQA {\bf People} category is exemplified with other contents having the similar outcomes.} \label{tab:resampling_ablation}
	\begin{tabular}{|c|c|c|c|c|}
		\hline
		Method&$\beta$ & PLCC & SROCC & RMSE \\ \cline{1-5}
		\multirow{5}{*}{{Random}}&${N}/{1e3}$ & 0.8836 & 0.8751 & 1.1561 \\ \cline{2-5}
		&${N}/{2e3}$& 0.8827 & 0.8755 & 1.1602 \\ \cline{2-5}
		&${N}/{5e3}$& 0.8864 & 0.8773 & 1.1426 \\ \cline{2-5}
		&${N}/{8e3}$& 0.8778 & 0.8689 & 1.1825 \\ \cline{2-5}
		&${N}/{1e4}$& 0.8796 & 0.8725 & 1.1744 \\ \cline{1-5}
		\multirow{5}{*}{{High-pass}~\cite{chen2017fast}} &${N}/{1e3}$ &  0.8896 & 0.8842 & 1.1274 \\ \cline{2-5}
		&${N}/{2e3}$& 0.8913 & 0.8841 & 1.1192 \\ \cline{2-5}
		&${N}/{5e3}$& 0.8856 & 0.8773 & 1.1467 \\ \cline{2-5}
		&${N}/{8e3}$& 0.8898 & 0.8835 & 1.1264 \\ \cline{2-5}
		&${N}/{1e4}$& 0.8874 & 0.8827 & 1.1380 \\ \cline{1-5}
	\end{tabular}
\end{table}
Random sampling introduces unstable performance from $\beta = {N}/{1e3}$ to ${N}/{1e4}$. This might be due to the reason that random sampled keypoints are not well covering the geometric structure, or frequency band sensitive to the perception. On the other hand, high-pass filtering retains the consistent performance across $\beta$s. It, to some extent, implies that as long as we can have the keypoints to accurately reflect the geometric structure (e.g., contours, edges), the number of keypoints can be sufficient smaller than the total points in the native point cloud.
The Haar-alike high-pass filter suggested in~\cite{chen2017fast} apparently meets this criteria. We expect that a better high-pass resampling would further improve the overall performance. But,  given the outstanding efficiency shown in Table~\ref{tab:resampling_ablation},   it has already demonstrated the model generalization to various resampling methodologies.

\begin{figure}[t]
	\centering
	\includegraphics[width=1\linewidth]{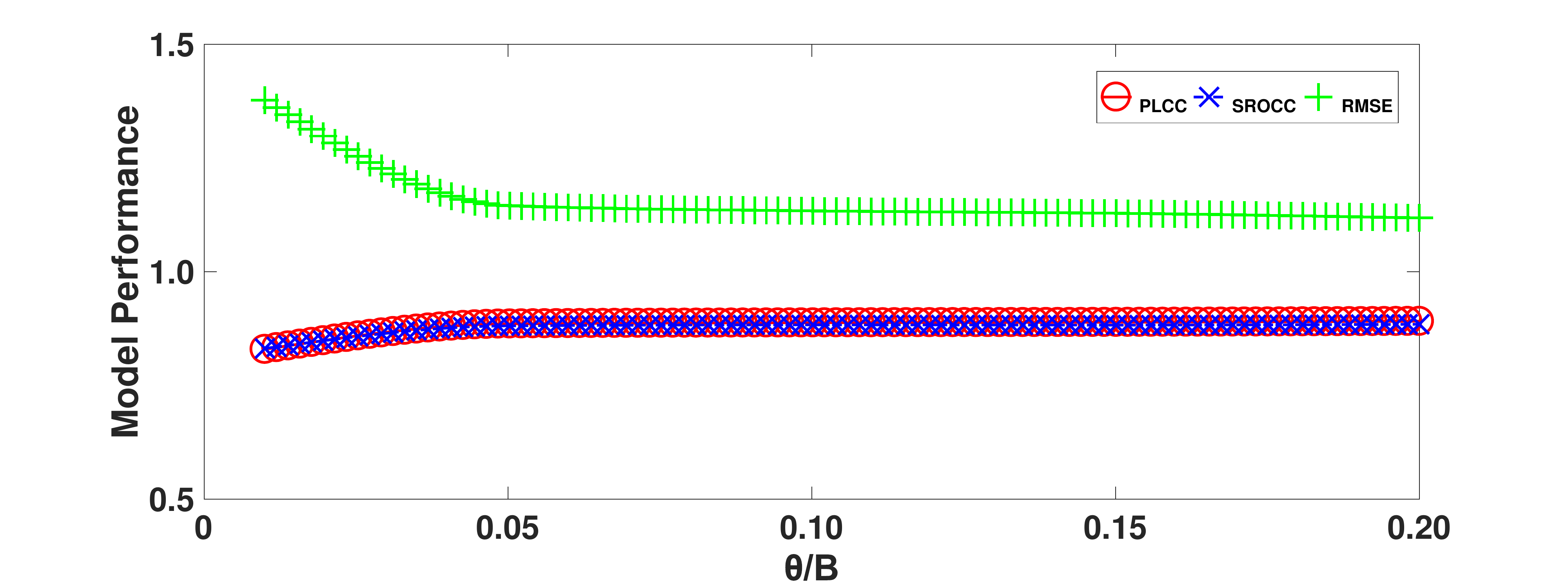}\\
	\caption{Model performance improves as $\theta$ increases.}
	\label{fig:graph_theta}
\end{figure}

\noindent{{\bf Neighbor Dimension.} Local graph is utilized as the basic unit for similarity derivation, which is derived from the clustered neighbors.  Given that $\tau$ and $\sigma$ are dependent on $\theta$, we first examine the impacts of different $\theta$ by setting $\frac{\theta}{B}$= 0.01, 0.05, 0.1, 0.15, and 0.2, as shown in Fig.~\ref{fig:graph_theta}. It reveals that the model performance can be quickly improved by enlarging the neighbor  dimension with larger $\theta$, and gets quite stable when  $\frac{\theta}{B}\geq 0.05$. This ensures the general applicability of the {\sf GraphSIM} as long as we give a reasonable $\theta$ bounded on the point cloud dimensional scale.}

\begin{figure}[t]
	\centering
	\includegraphics[width=1\linewidth]{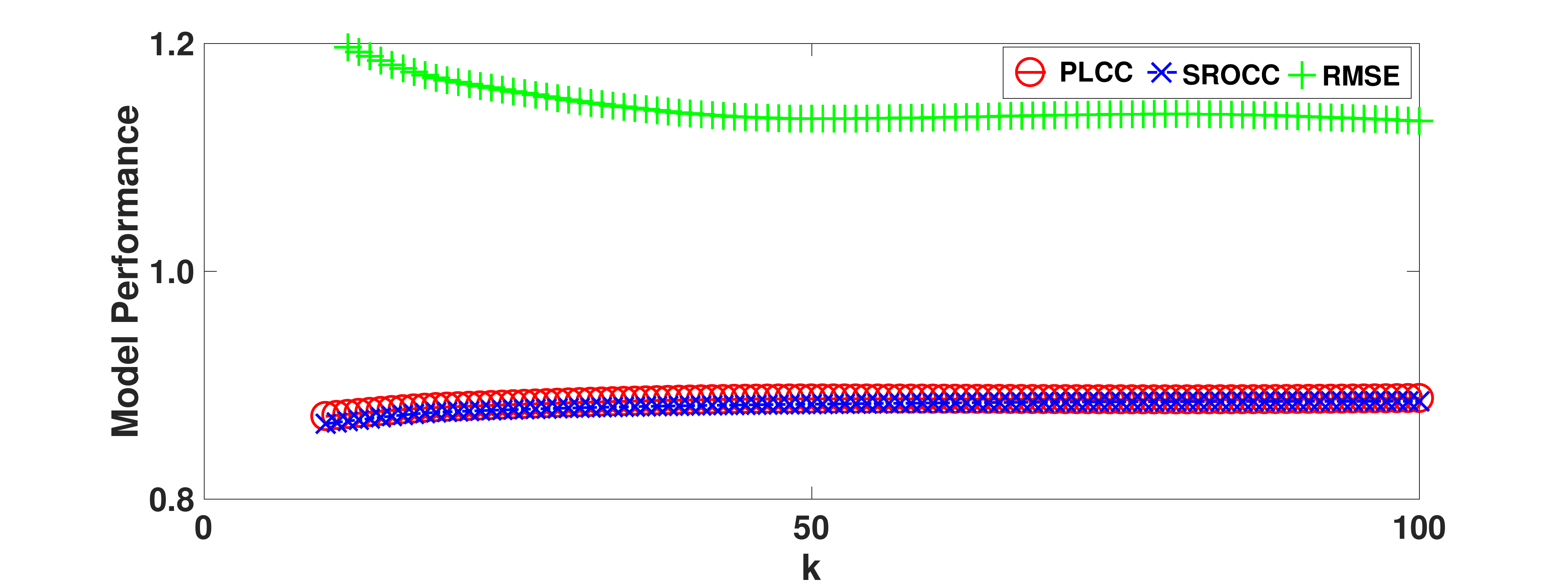}\\
	\caption{Model performance changes as $k$ increases.}
	\label{fig:tau}
\end{figure}

\noindent{{\bf Graph Scale.} $\tau$ is used as threshold to cluster neighbors into the same graph following \eqref{adjancy matrix}. Given that $\tau$ are dependent on the largest distance of the $k$-th nearest neighbor, we examine the impacts of different $k$ by setting ${k}$=10, 20, 50, 80, 100, as shown in Fig.~\ref{fig:tau}. Model performance can be gradually improved while $k$ increases. But it begins to stable when  $k>50$.  This also ensures the general applicability of the {\sf GraphSIM} with reasonable $k$ bounded on the point cloud graph scale.}

\subsection{Pooling Strategy}

 In Sec. \ref{sec:GraphSIM}, we first pool three feature-based similarities (e.g., $m_g$, $\mu_g$ and $c_g$) using {\it multiplication}, and then fuse color channels (e.g., R, G, B) using {\it average pooling}. Note that multiplication pooling is also used in SSIM index, and averaged pooling is widely used for overall PSNR derivation.

This part has attempted to examine different pooling methods in {\sf GraphSIM} for in-depth understanding of its capacity. We define the pooling method  ${P1}$ for feature similarity fusion under the same color channel, and ${P2}$ for the pooling across all color channels. Both ${P1}$ and ${P2}$ can adopt {\it multiplication} ($\mathbf{M}$) or {\it averaging} ($\mathbf{AVE}$). It then leads to four different combinations $C = [P1, P2]$, e.g., $C1=[\mathbf{
AVE}, \mathbf{AVE}]$, $C2=[\mathbf{M}, \mathbf{AVE}]$, $C3=[\mathbf{AVE}, \mathbf{M}]$ and $C4=[\mathbf{M}, \mathbf{M}]$. For $P1$, we distribute the same weighting factors, 1:1:1, for three features, while apply the 6:1:1 weighting factors for ${P2}$ assuming the GCM color model.

We use all samples in {\bf People} cateogy of SJTU-PCQA to study different pooling methods, with results shown in Table \ref{tab:pooling method}. Note that SROCC is relatively consistent for different pooling combinations. This is because both $\mathbf{M}$ and $\mathbf{AVE}$ do not change the monotonicity of test samples. On the contrary, PLCC and RMSE show obvious degradation when applying the $C4$, while $C1$ offers the best quantitative result. We believe that $\mathbf{M}$ aggravates the prediction error when fusing multiple feature-based similarities together.
 Assuming there are two samples, e.g., A and B, with respective MOSs as 9 and 8.
 We extract two features, e.g., $f_1$ and $f_2$, that will be utilized to derive individual similarity, e.g., ${\sf SIM}_{f_1}$, ${\sf SIM}_{f_2}$, for final objective index. For sample A,   ${\sf SIM}_{f_1}$ = ${\sf SIM}_{f_2}$ = 0.9; while for sample B, ${\sf SIM}_{f_1}$ = ${\sf SIM}_{f_2}$ = 0.8. Final objective scores for sample A is 0.81, and is 0.64 for sample B. If we use $\mathbf{AVE}$ instead, final scores are 0.9 and 0.8, showing higher correlation with MOS data. Alternatively, if one feature has high similarity, e.g., 0.9, while another one has very low similarity, e.g., 0.1, the results after performing the $\mathbf{M}$ or $\mathbf{AVE}$ will be much more different (e.g., $0.9*0.1=0.09$ vs. $(0.9+0.1)/2=0.5 $). Thus, without resorting for complex weighting factors for each individual feature (e.g., MOS data fitting), $\mathbf{AVE}$ is more reliable and robust than $\mathbf{M}$ for PLCC and RMSE measurement.

\begin{table}[t]
	\centering
	\caption{Model performance with various pooling methods.} \label{tab:pooling method}
	\begin{tabular}{|c|c|c|c|}
		\hline
		Method & PLCC & SROCC & RMSE \\ \cline{1-4}
		C1 & 0.8966 & 0.8938 & 1.0934 \\ \cline{1-4}
		C2 &  0.8896 & 0.8842 & 1.1274  \\ \cline{1-4}
		C3 & 0.8850 & 0.8780 & 1.1495 \\ \cline{1-4}
        C4 & 0.7827 & 0.8577 & 1.5364 \\ \cline{1-4}
	\end{tabular}
\end{table}

\subsection{Signal Type}\label{sec:signal_type}
Earlier studies assume the color attributes for similarity derivation. In practice, many applications process the uncolored point clouds (e.g., ModelNet40 \cite{wu20153d} samples). For instance, most uncolored point clouds are often used in computer vision tasks such as point cloud generation~\cite{achlioptas2018learning,groueix2018papier}. Therefore, this part of study discusses another two types of signal attributes, e.g., coordinate (G1) and normal (G2), without any color information. For G1, we directly use the 3D space coordinates to represent the signal attributes; For G2, we use the normal vector calculated by the principal components analysis (PCA) \cite{rusu2010semantic}. Besides, we further test the performance of {\sf GraphSIM} using the mixture of attributes, e.g., ``color+coordinate'' (M1), ``color+normal (M2)''. Note that the color signal used here is in GCM color space and the averaging (AVE) pooling strategy is applied for both channel-wise and attribute-wise aggregation. The experimental results are given in Table \ref{tab:signal_type}.
\begin{table}[t]
	\centering
	\caption{Model performance in terms of various types of signal attributes.} \label{tab:signal_type}
	\begin{tabular}{|c|c|c|c|}
		\hline
		Signal & PLCC & SROCC & RMSE \\ \cline{1-4}
		G1 & 0.8822 & 0.7864 & 1.1573 \\ \cline{1-4}
		G2 & 0.8836 & 0.8028 & 1.1557 \\ \cline{1-4}
		M1 & 0.9381 & 0.9228 & 0.8547 \\ \cline{1-4}
        M2 & 0.9374 & 0.9271 & 0.8596 \\ \cline{1-4}
	\end{tabular}
\end{table}

The {\sf GraphSIM} consistently present competitive performance for various signal attributes. For normal attributes (G2), it has the PLCC, SROCC, RMSE at respective 0.8836, 0.8028 and 1.1557. For G1, its SROCC is slightly degraded, e.g., $<0.80$. This is because,  when using the coordinates as the graph signal attribute,  $\mathbf{W}$ and [$f(\vec{X}_j)-f(\vec{s}_k)$] have complete opposite monotony with respect to the Euclidean distance of points. Note that $\mathbf{W}$ is an exponential function of point distance. Thus the impact of [$f(\vec{X}_j)-f(\vec{s}_k)$] is mostly
concealed. Moreover, it provides even better efficiency when having the mixed attributes. Note that the PLCCs of M1 and M2 are both above 0.93, and SROCCs are both above 0.92. All of these studies have clearly demonstrated the scalability, robustness of  the {\sf GraphSIM} for different signals.

\subsection{Graph Type}
we only use geometrical coordinates to construct local graphs, and then augment color attribute as graph signal to evaluate superimposed geometry and attribute distortions. As also report in Sec. \ref{sec:signal_type}, a mixture of attribute signal can further improve the model performance (see Table 10). Therefore, this section explores the possibility to include additional color information for constructing the local graph. For simplicity, we treat geometry and attribute components equally important, and apply the following equation to formulate graph edge weight
\begin{equation}\label{mixed_adjancy matrix}
\mathbf{W}_{\vec{X}_i,\vec{X}_j}^{'}=\left\{
\begin{aligned}
\frac{e^{-\frac{\|\vec{X}^O_{i}-\vec{X}^O_{j}\|^{2}_{2}}{\sigma_1^2}}+e^{-\frac{\|\vec{X}^I_{i}-\vec{X}^I_{j}\|^{2}_{2}}{\sigma_2^2}}}{2}&,\quad if \quad D\leq\tau;\\
0&,\quad else,
\end{aligned}
\right.
\end{equation}
with $\|\vec{X}^O_{i}-\vec{X}^O_{j}\|^{2}_{2}=D$, the coordinate and color normalized to [0, 1] for processing

We use ``RedandBlack'' and ``LongDress'' as test sequences, with results shown in Table \ref{tab:mixedgraph_ablation}. Following the definition in in Sec. \ref{sec:signal_type}, we test different attributes, including the color, G1 and M1 on constructed graphs.
\begin{table}[]
	\centering
	\caption{Model performance with different graph types: ``Redandblack'' and ``longDress'' of SJTU-PCQA {\bf People} category are exemplified with other contents having the similar outcomes.} \label{tab:mixedgraph_ablation}
	\begin{tabular}{|c|c|c|c|c|}
		\hline
		Graph Type&Signal Type & PLCC & SROCC & RMSE \\ \cline{1-5}
		\multirow{3}{*}{$\mathbf{W}$}&coordinate & 0.8763 & 0.7877 & 1.1399 \\ \cline{2-5}
		&color& 0.9063 & 0.8957 & 0.9998 \\ \cline{2-5}
		&mixed& 0.9417 & 0.9213 & 0.7961 \\ \cline{1-5}
		\multirow{3}{*}{{$\mathbf{W}^{'}$}} &coordinate &  0.7819 & 0.7754 & 1.4750 \\ \cline{2-5}
		&color& 0.8951 & 0.8952 & 1.0550 \\ \cline{2-5}
		&mixed& 0.8559 & 0.8577 & 1.2236 \\ \cline{1-5}
	\end{tabular}
\end{table}

Table \ref{tab:mixedgraph_ablation} reports that geometry-only $\mathbf{W}$ in \ref{adjancy matrix}  provides better performance than the $\mathbf{W}^{'}$ using both geometry and color attribute. for various signal types. This shows that there is no clear advantage to mix geometry and attribute information for graph construction.

\section{Conclusion}\label{sec:conclusion}

Point cloud techniques have advanced fast in recent years for virtual reality, telepresence, etc. However, it still lacks an efficient objective quality metrics that can accurately predict human perception, and can be embedded into the system for performance optimization. Existing point-wise distance-based metrics used in MPEG point cloud compression standards~\cite{schwarz2018emerging} are not only instable across contents and distortions, but also can not well reflect the perceptual sensation of the HVS.

Thus, we have developed the {\sf GraphSIM} to approach this problem by jointly considering the geometry and color distortions. It includes the point cloud resampling to extract keypoints (e.g., contours, edges) at high spatial-frequency that are more sensitive to the perception, followed by constructing the local graphs centered at extracted keypoints; and color gradient aggregation of each graph for final similarity index pooling across color channels and all graphs. Our {\sf GraphSIM} has demonstrated consistent, reliable correlation with the subjective MOSs upon two independent point cloud quality assessment datasets, presenting noticeable gains over the state-of-the-art metrics adopted in MPEG point cloud reference software.  GraphSIM parameters are either fixed constants, or directly dependent on the input signal (e.g., color space, bounding box scale, etc), making it fairly easy for system implementation. Ablation studies have further supported the model generalization by examining its key modules and model parameters.

There are several interesting avenues for future exploration. For example, how to better extend {\sf GraphSIM} for geometry point cloud  (i.e., without color attributes) is worth for deep investigation. Applying the {\sf GraphSIM} into MPEG point cloud compression technologies to quantitatively optimize the rate-distortion efficiency is another practical and attractive topic.

\section{Acknowledgment}
We would like to thank anonymous reviewers for their constructive comments to improve this manuscript. In the meantime, we really appreciate the efforts devoted in~\cite{javaheri2019point} for developing the IPRC point cloud assessment database.

\ifCLASSOPTIONcaptionsoff
  \newpage
\fi


\bibliography{sample-base}
\bibliographystyle{IEEEtran}

\ifCLASSOPTIONcaptionsoff
\newpage
\fi
\vspace{-10 mm}
\end{document}